\documentclass[%
 aip,
 sd,%
 amsmath,amssymb,
 reprint,%
]{revtex4-1}
\usepackage[version=3]{mhchem}
\usepackage[left=1.5cm, right=1.5cm, top=1.785cm, bottom=2.0cm]{geometry}
\usepackage{balance}
\usepackage{times,mathptmx}
\usepackage{float}
\usepackage[english]{babel}
\usepackage{natbib}
\usepackage{gensymb}
\usepackage{hyphenat}
\usepackage{siunitx}
\usepackage{array}
\usepackage[usenames,dvipsnames]{xcolor}
\usepackage{hyperref}
\usepackage{graphicx}
\usepackage{dcolumn}
\usepackage{bm}
\usepackage[normalem]{ulem}

\newlength{\figwidth}
\newlength{\figwidthwide}
\let\orgautoref\autoref
\providecommand{\Autoref}{%
  \def\equationautorefname{Equation}%
  \def\figureautorefname{Figure}%
  \def\subfigureautorefname{Figure}%
  \orgautoref}
\renewcommand{\autoref}{%
  \def\equationautorefname{Eq.}%
  \def\figureautorefname{Fig.}%
  \def\subfigureautorefname{Fig.}%
  \orgautoref}

\begin{document}

\preprint{AIP/123-QED}

\title{Structure determination of the tetracene dimer in helium nanodroplets using femtosecond strong-field ionization}

\author{Constant Schouder}
\affiliation{Department of Physics and Astronomy, Aarhus University, Denmark}
\author{Adam S. Chatterley}%
\affiliation{Department of Chemistry, Aarhus University, Denmark}

\author{Florent Calvo}
\affiliation{Universit\'{e} Grenoble Alpes, LIPHY, F-38000 Grenoble, France}
\author{Lars Christiansen}
\affiliation{Department of Chemistry, Aarhus University, Denmark}
\author{Henrik Stapelfeldt}
\affiliation{Department of Chemistry, Aarhus University, Denmark}

\date{\today}

\begin{abstract}
Dimers of tetracene molecules are formed inside helium nanodroplets and identified through covariance analysis of the emission directions of kinetic tetracene cations stemming from femtosecond laser-induced Coulomb explosion. Next, the dimers are aligned in either one or three dimensions under field-free conditions by a nonresonant, moderately intense laser pulse. The experimental angular covariance maps of the tetracene ions are compared to calculated covariance maps for seven different dimer conformations and found to be consistent with four of these.  Additional measurements of the alignment-dependent strong-field ionization yield of the dimer narrows the possible conformations down to either a slipped-parallel or parallel-slightly-rotated structure. According to our quantum chemistry calculations, these are the two most stable gas-phase conformations of the dimer and one of them is favorable for singlet fission.
\end{abstract}

\maketitle


\section{\label{sec:Introduction}Introduction}

Noncovalent interactions between aromatic molecules are crucial for many areas, such as molecular recognition, structure of macromolecules and organic solar cells.~\cite{hobza_world_2006,a.ikkanda_exploiting_2016,becucci_high-resolution_2016,smith_singlet_2010} At the most fundamental level, the interaction involves two aromatic molecules. This has been the subject of a large numbers of studies, often with a particular focus on determining the structure of the dimers. Experimentally, the main technique to form dimers is supersonic expansion of a molecular gas seeded in a carrier gas of rare gas atoms into vacuum. Combining the resulting molecular beams with various types of high-resolution spectroscopy, including microwave, infrared, and UV spectroscopy~\cite{becucci_high-resolution_2016,nesbitt_high-resolution_1988,moazzen-ahmadi_spectroscopy_2013} as well as rotational coherence spectroscopy~\cite{felker_rotational_1992,riehn_high-resolution_2002} --- a technique based on pairs of femtosecond or picosecond pulses --- the rotational constants can be extracted. Upon comparison with results from theoretical modelling, information about the conformations of a range of different dimers have been obtained.~\cite{doi:10.1146/annurev.physchem.49.1.481, becucci_high-resolution_2016} Examples include the dimers of benzene,~\cite{doi:10.1063/1.465035} fluorene,~\cite{doi:10.1063/1.466382} benzonitrile,~\cite{doi:10.1063/1.452252,SCHMITT2006234} phenol,~\cite{doi:10.1002/cphc.200500670} and anisole.~\cite{doi:10.1021/jp903236z}

An alternative experimental method is to form molecular dimers or larger oligomers inside helium nanodroplets.\cite{choi_infrared_2006,yang_helium_2012} This technique makes it possible to create aggregates of much larger molecules, for instance of fullerenes~\cite{jaksch_electron_2009} and polycyclic aromatic hydrocarbons,\cite{wewer_molecular_2003,roden_vibronic_2011,birer_dimer_2015} than what is typically possible in molecular beams from standard supersonic expansions. Furthermore, the variety of heterogenous aggregates goes beyond the normal reach of the gas phase.\cite{choi_infrared_2006,yang_helium_2012}  Structure characterization has mainly been obtained by IR spectroscopy~\cite{nauta_nonequilibrium_1999,sulaiman_infrared_2017,verma_infrared_2019} although for complexes of larger molecules this becomes very challenging, as the density of states is too high to clearly resolve peaks in the spectra.

Recently, we introduced an alternative method for structure determination of dimers created inside He droplets, namely Coulomb explosion induced by an intense femtosecond (fs) laser pulse and recording of the emission direction of the fragment ions including identification of their angular correlations,~\cite{pickering_alignment_2018,pickering_femtosecond_2018} implemented through covariance analysis.~\cite{hansen_control_2012,slater_covariance_2014,frasinski_covariance_2016} Crucial to the structure determination was that the dimers had a well-defined spatial orientation prior to the Coulomb explosion event, in practice obtained by laser-induced alignment with a moderately intense laser pulse.\cite{stapelfeldt_colloquium:_2003,fleischer_molecular_2012} While the Coulomb explosion method may not match the level of structural accuracy possible with high resolution spectroscopy, at least not for comparatively small molecules, it distinguishes itself by the fact that the structure is captured within the time scale of the pulse duration, i.e. in less than 50 fs. As such, this technique holds the potential for imaging the structure of dimers as they undergo rapid structural change, for example due to excimer formation. The purpose of the current manuscript is to show that the Coulomb explosion method can also be used to obtain structure information about dimers composed of molecules much larger than the carbon disulfide and carbonyl sulfide molecules studied so far.\cite{pickering_alignment_2018,pickering_femtosecond_2018} Here we explore the dimer of tetracene (Tc), a polycyclic aromatic hydrocarbon (PAH) composed of four linearly fused benzene rings. We demonstrate that a covariance map analysis of the angular distributions of fragments from fs laser-induced Coulomb explosion supplemented by measurement of alignment-dependent ion yields allow us to identify the tetracene conformation as a face-to-face structure with the tetracene monomers either slightly displaced or slightly rotated. This identification relies on comparison of the experimental covariance maps to simulated covariance maps for a range of plausible conformations.

Our motivation for exploring the structure of the dimer is threefold. Firstly, PAHs are know to be relatively abundant in the interstellar
medium,~\cite{doi:10.1146/annurev.astro.46.060407.145211} and laboratory experiments are required to deduce the signatures
that astronomers will need to hunt for. Secondly, PAH oligomers of similar size such as the pyrene dimer
have a possibly key role in the formation of soot,~\cite{sabbah10} and
it is important to characterize their structure and bonding with
accuracy. Thirdly, noncovalently bonded ensembles of tetracene and
other acenes can undergo singlet fission,~\cite{smith_singlet_2010,smith_recent_2013}
a phenomenon with major implications for solar energy harvesting
where a singlet excited state decays into two triplet states localized
on two separate monomers. The process produces two excitations from a
single photon, which in principle provides a means to overcome the
Shockley-Queisser $33\%$ efficiency limit inherent to single
excitation photovoltaic systems. Typically, singlet fission is studied in crystals
and thin films, which are excellent for emulating photovoltaic
devices, but due to their extended nature are less well suited for
investigating the basic photophysical mechanisms, and so fundamental
understandings of singlet fission have somewhat lagged behind technological
developments. Solution studies of carefully synthesized covalent dimer
systems with only two acene units provide clearer insight into the
fundamental dynamics, with the caveat that the covalent linkage may
interfere with the electronic structure, and solvent and temperature
effects blur
spectra.~\cite{Zirzlmeier5325,doi:10.1021/acs.jpca.6b00988} Helium nanodroplets allow formation of the pristine dimers without the
need for an extended system or any covalent bond. Structural determination is essential as the relative configuration of the monomer in a polyacene dimer is crucial for the singlet fission process: the $\pi$ systems must overlap, but there must also be a on offset between the two units for singlet fission to be an allowed process.~\cite{smith_singlet_2010,smith_recent_2013} The conformations observed in the present work are favorable for
singlet fission, paving the way for future time-resolved studies singlet fission processes
in controlled environments.


\section{Experimental Setup}

\label{Sec:ExpSetup}

\begin{figure}[h]
\centering
\includegraphics[width=\columnwidth]{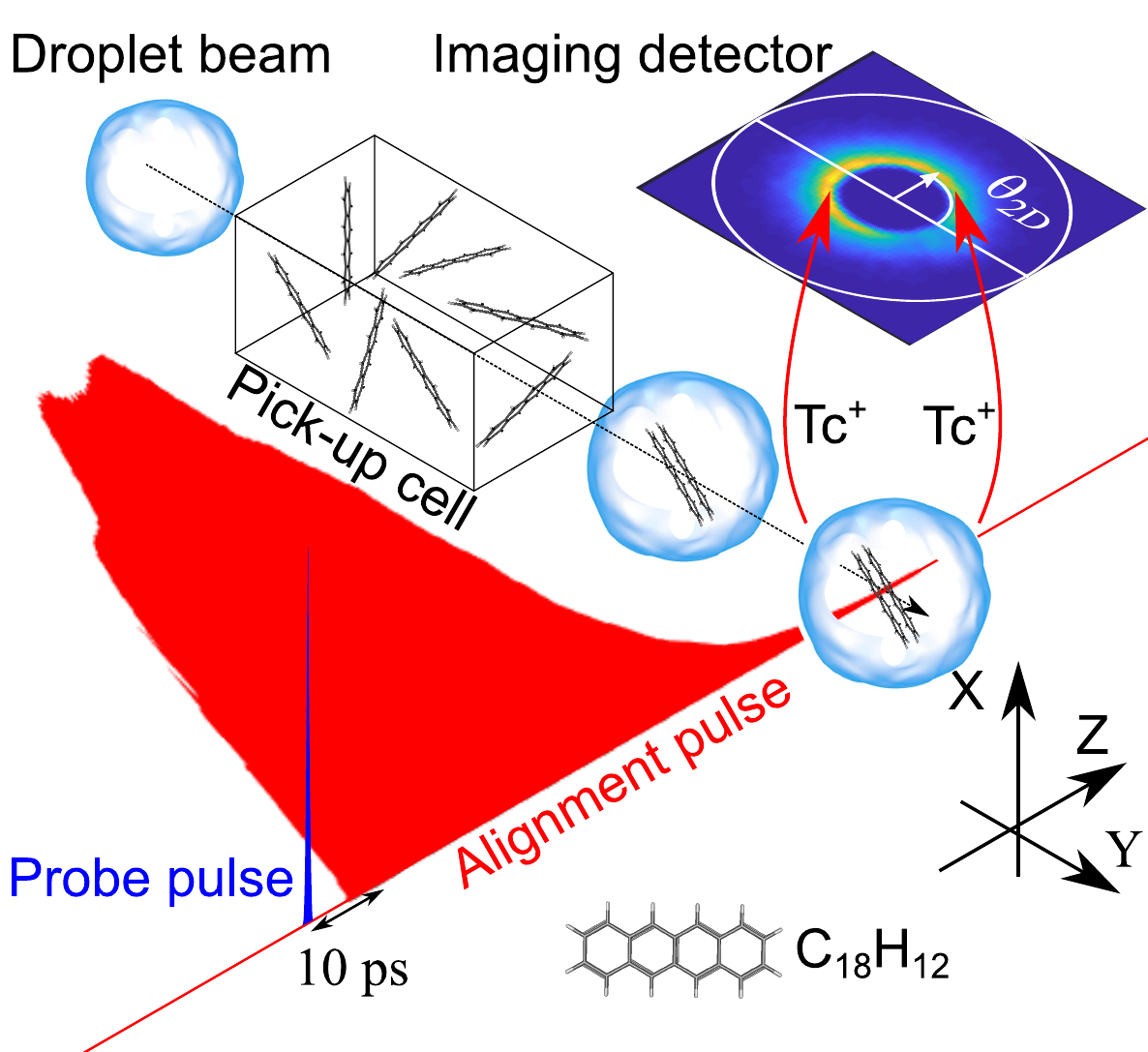}
\caption{Schematic of the key elements in the experiment. Tc dimers are first formed inside He droplets, then aligned by a truncated nonresonant alignment laser pulse (red). 10 ps after truncation of the alignment pulse, the Tc dimers are doubly ionized by a short intense probe pulse (blue) to induce Coulomb explosion and the velocities of the two Tc$^+$ ions are measured using velocity map imaging. In all measurements the probe pulse was linearly polarized along the X axis (perpendicular to the detector plane) whereas the alignment pulse was either linearly polarized along the X axis or the Y axis (depicted here), or elliptically polarized with the major polarization along the X axis or the Y axis.}
\label{Fig:ExpSetup}
\end{figure}

The experimental setup has been described in detail before~\cite{doi:10.1063/1.4983703} and only important aspects are pointed out here. A schematic of the setup is shown in \autoref{Fig:ExpSetup}. A continuous beam of He droplets is formed by expanding He gas through a \SI{5}{\mu m} nozzle, cooled to \SI{12}{K}, into vacuum. The backing pressure is 25 bar leading to droplets consisting on average of \SI{10000} He atoms.~\cite{doi:10.1002/anie.200300611} The droplets then pass through a pickup cell containing tetracene vapor obtained by resistively heating a sample of solid tetracene. The probability for a droplet to pick up one or two \ce{Tc} molecules depends on the partial pressure of tetracene in the cell~\cite{doi:10.1146/annurev.physchem.49.1.1} defined by its temperature. As discussed in \autoref{sec:results-coulomb}, this allows us to control the formation of \ce{Tc} dimers inside the He droplets.

Hereafter, the doped He droplet beam enters the target region where it is crossed, at right angles, by two collinear, focused, pulsed laser beams. The pulses in the first beam are used to induce alignment of the \ce{Tc} dimers (and monomers) in the He droplets. They have an asymmetric temporal shape rising to a peak in $\sim$120 ps and turning-off in $\sim$10 ps --- see sketch in \autoref{Fig:ExpSetup} and measured intensity profile in \autoref{Monomeryield} and \autoref{Dimeryield} --- obtained by spectrally truncating the uncompressed pulses from an amplified Ti:Sapphire laser system.~\cite{doi:10.1063/1.5028359} Their central wavelength is \SI{800}{\nano\metre}, the focal spot size, $\omega_0$, is \SI{65}{\mu m}, and the peak intensity $\sim$\SI{0.16}{TW/cm^2}. The pulses in the second beam are used to identify the formation of \ce{Tc} dimers and measure their alignment. As detailed in \autoref{sec:results-coulomb}, this relies on ionization of the \ce{Tc} dimers. These probe pulses (35 fs long, $\lambda_\text{central}$ = \SI{400}{\nano\metre}) are created by second harmonic generation of the compressed output from the Ti:Sapphire laser system in a 50 $\mu$m thick BBO crystal. The focal spot size, $\omega_0$, is \SI{50}{\mu m} and the intensity is varied from 0.6 to \SI{9}{TW/cm^2}. The repetition rate of the laser pulses is 1 kHz.

The \ce{Tc}$^+$ ions, created by the probe pulse, are projected by a velocity-map imaging spectrometer onto a 2D imaging detector. The detector consists of two microchannel plates backed by a P47 phosphor screen, whose images are recorded on a CCD camera. The CCD camera is read out every 10 laser shots, i.e. at a 100 Hz rate, and such an image is termed a frame. Ion images as the ones shown in \autoref{fig:ion-images-covariance} typically consist of 10~000 frames.

\section{Molecular modelling of the tetracene dimer conformation}
\label{sec:qchem}

The conformations of the tetracene dimer were independently explored
by molecular modeling and quantum chemical methods in order to identify
plausible candidates for interpreting the measurements.

\begin{figure}
  \centering
  \includegraphics[width=7cm]{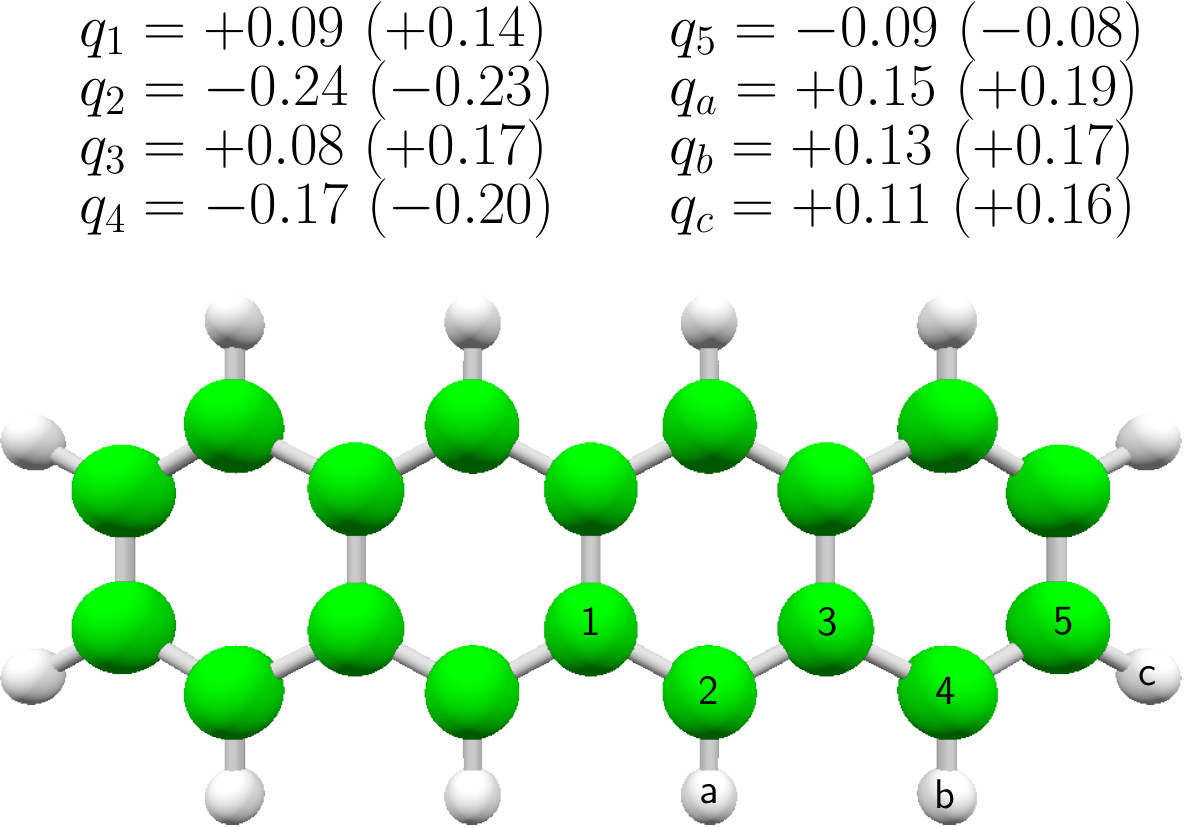}
  \caption{Partial charges on carbon and hydrogen atoms on the
    tetracene monomer neutral and cation in brackets, as used in the force field exploration of dimer
    conformations and in the simulations. All charges are expressed in units of the electron charge.}
\label{Fig:tetracenemonomer}
\end{figure}

The potential energy landscape was first explored using a simple
quantum mechanical model previously developed to simulate the sticking
between PAH molecules under astrophysical
conditions.~\cite{rapacioli06} Briefly, the model consists of an additive
potential with intramolecular contributions $V_{\rm intra}$ for each
tetracene molecule, and a pairwise force field $V_{\rm inter}$
describing the non-covalent interactions between the two flexible
molecules. Here $V_{\rm intra}$ is based on an earlier tight-binding
model of Van-Oanh and coworkers~\cite{vanoanh02} while $V_{\rm inter}$
is a simple sum of Lennard-Jones (LJ) and Coulomb terms acting between
the atomic positions that also carry partial charges representing the
multipolar distribution. For the LJ potential we employed two
parameter sets either from the OPLS library~\cite{opls} or published
earlier by van de Waal in the context of hydrocarbon clusters.~\cite{vdw83} The partial charges on tetracene were evaluated
using the fluctuating charges method,~\cite{mortier06} as proposed by
Rapacioli and coworkers~\cite{rapacioli05} who adjusted its parameters
so that it mimics the RESP procedure often used to extract charges
from DFT calculations. The charges obtained for the tetracene monomer
are shown in \autoref{Fig:tetracenemonomer}.

Our initial scanning procedure consisted of a large amplitude Monte
Carlo exploration of the possible conformations of the dimer,
followed by systematic local optimization using this flexible
potential. The optimized geometries were then refined directly with
quantum chemical methods, employing here again DFT with functionals
that account for long-range (noncovalent) forces that are essential
for the present system. The two functionals B97-1 and wB97xD were thus
employed with the two basis sets 6-311G(d,p) and TZVP. Basis set
superposition errors were accounted for using the standard
counterpoise method, and from the equilibrium geometries the harmonic
zero-point energies were also evaluated. All quantum chemical
calculations were performed using Gaussian09.~\cite{G09}

Our exploration lead to only two locally stable conformations, with
the two molecular planes parallel to each other, the main axes of the
molecules being themselves either parallel as well or forming an angle
of about 25$^\circ$. In the former case, the molecules are not
superimposed on each other but shifted in order to maximize van der
Waals interactions (as in graphite). Consistently with standard
terminology,~\cite{rapacioli05} the resulting conformations, numbered
as 1 and 2 in \autoref{Fig:Covariance}, are referred to as parallel
displaced or rotated, respectively. The other conformations shown in
\autoref{Fig:Covariance}, numbered 3--7, are not locally stable with
any of the methods used and relax into either of the two parallel
conformers.

The relative energies of the two conformers are compared in \autoref{table}, and the binding energy of the most stable one is
provided as well.
\begin{table}[htb]
   \begin{tabular}{c|r|r}
     \hline\hline
     Method & Parallel displaced & Rotated \\
     \hline
     TB+LJ (vdW) & +4.9 (+6.5) & {\bf 582.0 (554.1)} \\
     TB+LJ (OPLS) & +6.8 (+8.1) & {\bf 437.5 (416.7)} \\
     B97-1/6-311G(d,p) & +3.3 (+4.6) & {\bf 156.1 (148.6)} \\
     B97-1/TZVP & {\bf 116.3 (100.4)} & +36.5 (+25.6) \\
     wB97xD/6-311G(d,p) & +26.6 (+28.1) & {\bf 893.4 (868.1)} \\
     wB97xD/TZVP & +19.9 (+16.7) & {\bf 730.7 (698.9)} \\
     \hline
     \hline
   \end{tabular}
   \caption{Binding energies and relative energies of the parallel
     displaced and rotated conformers of the tetracene dimer, as
     obtained from a simple quantum mechanical force field (TB+LJ) with
     two sets of LJ parameters, or from density-functional theory
     minimizations with two functionals and two basis sets and after
     correcting for basis set superposition error. Absolute numbers
     (also in bold face) are the binding energies obtained for the most
     stable conformer, numbers with a plus sign indicate the relative
     difference of the less stable conformer. The values in parentheses
     include the harmonic zero-point energy corrections. All values are
     given in meV.}
   \label{table}
\end{table}
From this table we find a significant spreading in the values of the
binding energy, which roughly varies from 200~meV for the B97-1 DFT
method to 800~meV for the wB97xD method, the empirical models yielding
values of about 500~meV in between these two extremes. However, the
energy difference between the conformers appears as a fraction of the
absolute binding energy whatever the level of calculation. From the
perspective of the quantum force field, the two conformers are nearly
isoenergetic, with the parallel displaced isomer being higher by less
than 10~meV. The DFT results generally predict the same ordering, but
with a slightly higher difference closer to 20--30~meV depending on
whether zero-point correction is included or not, still quite
small. Only the B97-1 functional with the TZVP basis set finds
otherwise that the rotated conformer should not be the most stable of
the two. At this stage, we cautiously conclude that two particular
conformations for the tetracene dimer are candidates for experimental
elucidation, both having the molecules parallel to each other but some
shift or rotation between their main symmetry axes.

\section{Results: Coulomb explosion}
\label{sec:results-coulomb}

\subsection{Identification of tetracene dimers}
\label{sec:identification}
\begin{figure}
\centering
\includegraphics[width=\columnwidth]{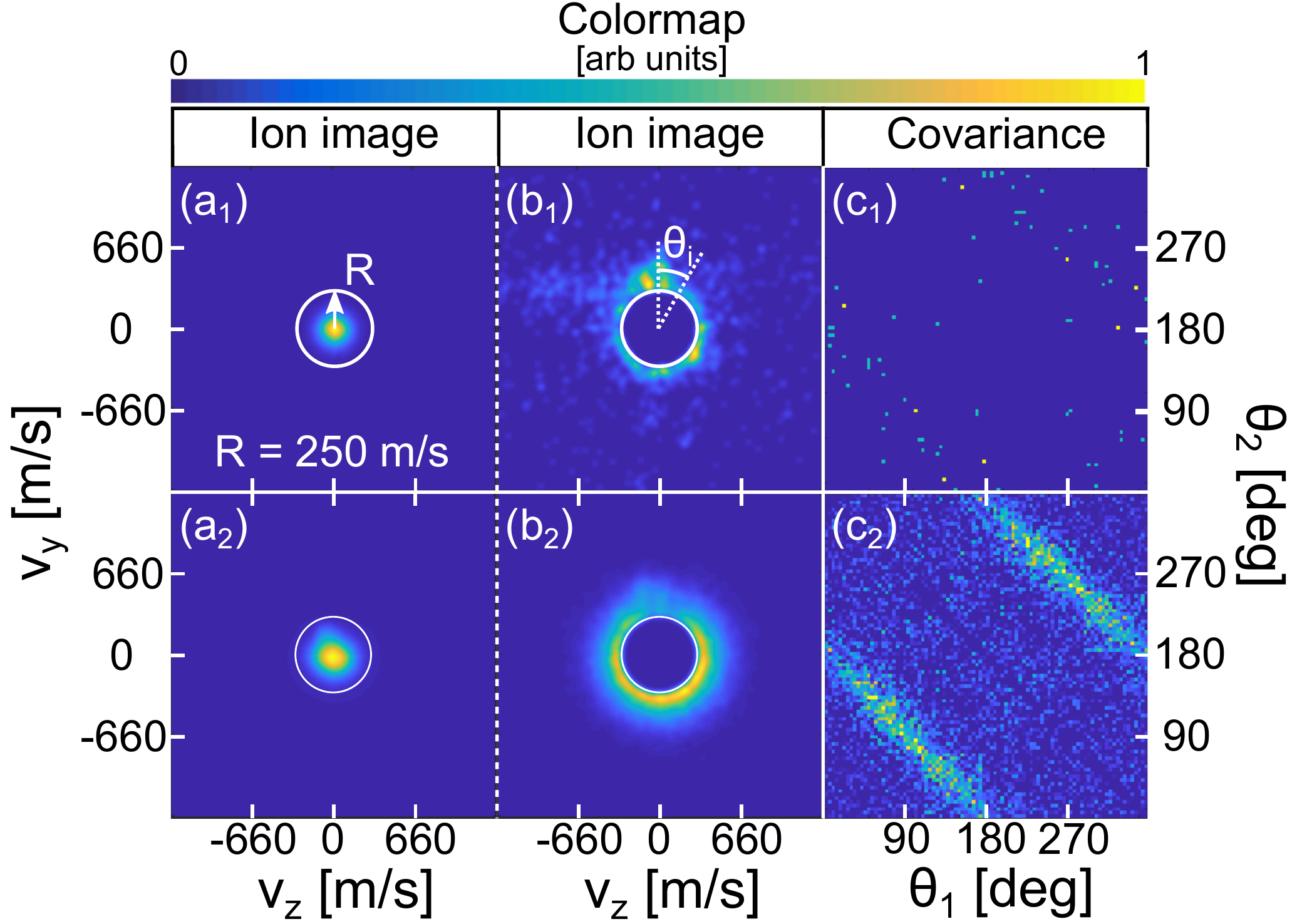}
\caption{(a$_1$) \ce{Tc^+} ion image recorded for a low partial pressure of tetracene in the pickup cell (monomer doping condition); (a2) \ce{Tc^+} ion image recorded for a higher partial pressure of tetracene in the pickup cell (dimer-doping condition); (b1)-(b2) Same image as (a1) and (a2) but with the center removed; (c1)-(c2) Corresponding angular covariance maps created from ions count outside the white circles. The ion images are obtained with without alignment, with the probe pulse at an intensity of 3 \si{TW/cm^2}}
\label{fig:ion-images-covariance}
\end{figure}

First, we show that it is possible to form and detect \ce{Tc} dimers in the He droplets. The strategy is the same as that recently applied to identify \ce{CS2} and \ce{OCS} dimers~\cite{pickering_alignment_2018,pickering_femtosecond_2018} and relies on detecting kinetic \ce{Tc^+} ions as a sign of dimers. In detail, if a droplet contains just one \ce{Tc} molecule and this molecule is ionized by the probe pulse, the resulting \ce{Tc^+} ion will have almost zero kinetic energy. By contrast, if a droplet contains a dimer and both of its monomers are singly ionized, the internal Coulomb repulsion of the \ce{Tc^+} ions will cause them to gain kinetic energy. \Autoref{fig:ion-images-covariance} (a$_1$) shows a \ce{Tc^+} ion image recorded with the probe pulse only. The ions are localized in the very center of the image with more than 98 percent of them having a velocity less than \SI{250}{m/s}. These low-velocity ions are ascribed as originating from the ionization of single-doped tetracene molecules.  The ionization potential of tetracene is 6.97 eV and the photon energy of the probe photons is 3.1 eV. Thus, we believe that ionization is the result of 3-photon absorption by the tetracene molecules.

\Autoref{fig:ion-images-covariance}(a$_2$) also shows a \ce{Tc^+} ion image recorded with the probe pulse only but for a higher partial pressure of the tetracene gas in the pickup cell. The image is still dominated by an intense signal in the center but now ions are also being detected at larger radii corresponding to higher velocities. This can be highlighted by cutting the center in the image. The images in \autoref{fig:ion-images-covariance}(b$_1$) and (b$_2$) are the same as the images in \autoref{fig:ion-images-covariance}(a$_1$) and (a$_2$), respectively, but with a central cut removing contributions of \ce{Tc^+} ions with a velocity lower than \SI{250}{m/s}. It is now clear that the image in \autoref{fig:ion-images-covariance}(b$_2$) contains a significant amount of ions away from the central part. We assign the high-velocity ions to ionization of both \ce{Tc} molecules in droplets doped with a dimer.

To substantiate this assignment, we determined if there are correlations between the emission directions of \ce{Tc^+} ions, implemented through covariance analysis. Let $X^{(i)}$ be the discrete random variable that denotes the number of ions detected at an angle $\theta_i$ with respect to the vertical center line, see \autoref{fig:ion-images-covariance}\,(b$_1$). Experimentally, the detected ions are binned into $M$ equal-size intervals over the 360{\degree} range and thus the angular distribution of the ions can be represented by the vector:
\begin{equation}
\mathbf{X} = \{X^{(1)},X^{(2)}, . . . X^{(M)}\}.
\end{equation}
As mentioned in \autoref{Sec:ExpSetup} the resulting ion images are averaged over a large number, $N$, of individual frames. In practice, the angular distribution is therefore given by the expectation value of {\bf X}:
\begin{equation}
\langle\mathbf{X}\rangle =  \{\frac{1}{N}\sum_{n=1}^{N}x_n^{(1)}, \frac{1}{N}\sum_{n=1}^{N}x^{(2)}_n, . . .,  \frac{1}{N}\sum_{n=1}^{N}x^{(M)}_n \}
\end{equation}
where $x^{(i)}_n$ is the outcome (number of ions) of the random variable $X^{(i)}_n$ related to the angle $\theta_i$ for the $n^{\rm th}$ frame acquired. The covariance can now be calculated by the standard expression:
\begin{equation}
\text{Cov}(\mathbf{X},\mathbf{X}) = \langle \mathbf{XX} \rangle
- \langle \mathbf{X} \rangle\langle \mathbf{X} \rangle.
\end{equation}
using the ions in the radial range outside of the annotated white circle. The result, displayed in \autoref{fig:ion-images-covariance}(c$_2$), is referred to as the angular covariance map. We used an angular bin size of 4 degrees which gives $M = 90$. The covariance map reveals two distinct diagonal lines centered at $\theta_2 = \theta_1 \pm 180{\degree}$. These lines show that the emission direction of a \ce{Tc^+} ion is correlated with the emission direction of another \ce{Tc^+} ion departing in the opposite direction. This strongly indicates that the ions originate from ionization of both \ce{Tc} molecules in dimer-containing droplets and subsequent fragmentation into a pair of \ce{Tc^+} ions. Therefore, we interpret the angular positions of the \ce{Tc^+} ion hits outside the white circle as a measure of the (projected) emission directions of the \ce{Tc^+} ions from dimers. Note that the angular covariance signal extends uniformly over 360{\degree}. This shows that the axis connecting the two \ce{Tc} monomers is randomly oriented at least in the plane defined by the detector. This is to be expected in the absence of an alignment pulse.  Also note that at the low partial pressure of tetracene, used for the image in \autoref{fig:ion-images-covariance}\,(a$_1$), the pronounced lines in the angular covariance map are no longer present, see \autoref{fig:ion-images-covariance}\,(c$_1$), indicating that there are essentially no dimers under these pickup conditions.

\subsection{Angular covariance maps for aligned tetracene dimers}
\label{sec:ang-cov-maps}

Next, we carried out experiments aiming at determining the conformation of the \ce{Tc} dimers. In the first set of measurements, described in this section, the dimers are aligned then double ionized with the probe pulse, and the emission direction of the \ce{Tc+} ions is recorded. We then calculated their angular covariance maps, which were proven to provide useful information about the dimer conformation in the cases of \ce{CS2} and \ce{OCS} dimers.~\cite{pickering_alignment_2018,pickering_femtosecond_2018}

Based on recent findings for similar-size molecular systems, we expect that the strongest degree of alignment occurs around the peak of the alignment pulse and that, upon truncation of the pulse, the degree of alignment lingers for 10--20 ps, thereby creating a time window where the alignment is sharp and the alignment pulse intensity reduced by several orders of magnitude.~\cite{Chatterley2019} It is crucial to synchronize the probe pulse to this window because \ce{Tc+} ions are fragmented when created in the presence of the alignment pulse. The time-dependent measurements of the \ce{Tc+} yields, presented in \autoref{Sec:Ionization anisotropy}, shows this effect explicitly. Consequently, the probe pulse is sent 10 ps after the peak of the alignment pulse as sketched in \autoref{Fig:ExpSetup}. In the previous experiments on \ce{CS2} and \ce{OCS} dimers, the probe pulse was sent at the peak of the alignment pulse, and no truncation was needed, because \ce{CS2^+} and \ce{OCS^+} ions can both survive the alignment field.

The experimental results, recorded for different polarization states of the alignment pulse, are presented in the second column of \autoref{Fig:Covariance}. The partial pressure of tetracene vapor was set to the same value as that used for the data in row (2) in \autoref{fig:ion-images-covariance}, i.e. under doping conditions where a significant number of the He droplets contain dimers. The \ce{Tc+} ion images are not shown, only the angular covariance maps created from the ions in the corresponding images originating from ionization of the dimers. In practice, these are the ions detected outside of a circle with the same diameter as the one shown in \autoref{fig:ion-images-covariance}\,(a$_2$). The intensity of the probe pulse was again \SI{3}{TW/cm^2} for all the angular covariance maps shown in \autoref{Fig:Covariance}

The different rows in \autoref{Fig:Covariance} are the results of different polarization states of the alignment pulse and thus different spatial alignments of the dimers. In row (a), the alignment pulse was linearly polarized perpendicular to the detector plane, i.e. along the X axis --- see \autoref{Fig:ExpSetup}. This induces 1D alignment with the most polarizable axis of the \ce{Tc} dimer confined along the X axis. The covariance map [\autoref{Fig:Covariance}(a$_\text{e}$)] contains two prominent stripes, centered at $\theta_2 = \theta_1 \pm 180{\degree}$, very similar to those observed with the probe pulse only [\autoref{fig:ion-images-covariance}\,(b2)]. This implies that the axis connecting the two \ce{Tc} monomers is randomly oriented perpendicular to the X axis. \Autoref{Fig:Covariance}(b$_\text{e}$) is also obtained for 1D aligned dimers but now the most polarizable axis, defined by the polarization direction of the alignment pulse, is confined along the Y axis, i.e. in the detector plane. It is seen that the covariance signal no longer extends over all angles but rather appears as two islands centered at (\SI{90}{\degree},\SI{270}{\degree}) and (\SI{270}{\degree},\SI{90}{\degree}), respectively. Panels (c$_\text{e}$) and (d$_\text{e}$) of \autoref{Fig:Covariance} were recorded with an elliptically polarized alignment pulse with an intensity ratio of 3:1 in order to induce 3D alignment~\cite{larsen_three_2000,nevo_laser-induced_2009-1} where the most polarizable axis of the dimer is confined along the major polarization axis (parallel to the X axis in panel c and to the Y axis in panel d) and the second most polarizable axis along the minor polarization axis (parallel to the Y axis in panel c and to the X axis in panel d). The covariance maps also show islands localized around (\SI{90}{\degree},\SI{270}{\degree}) and (\SI{270}{\degree},\SI{90}{\degree}). Finally, the dimers were aligned with a circularly polarized alignment pulse, which confines the plane of the dimer to the polarization (X,Y) plane, but leaves it free to rotate within this plane. Again, the covariance signals are two islands localized around (\SI{90}{\degree},\SI{270}{\degree}) and (\SI{270}{\degree},\SI{90}{\degree}).

\subsection{Comparison of experimental covariance maps to simulated covariance maps}
\label{sec:comparison-ang-cov-maps}

\begin{figure*}
        \centering
    \includegraphics[scale = .55]{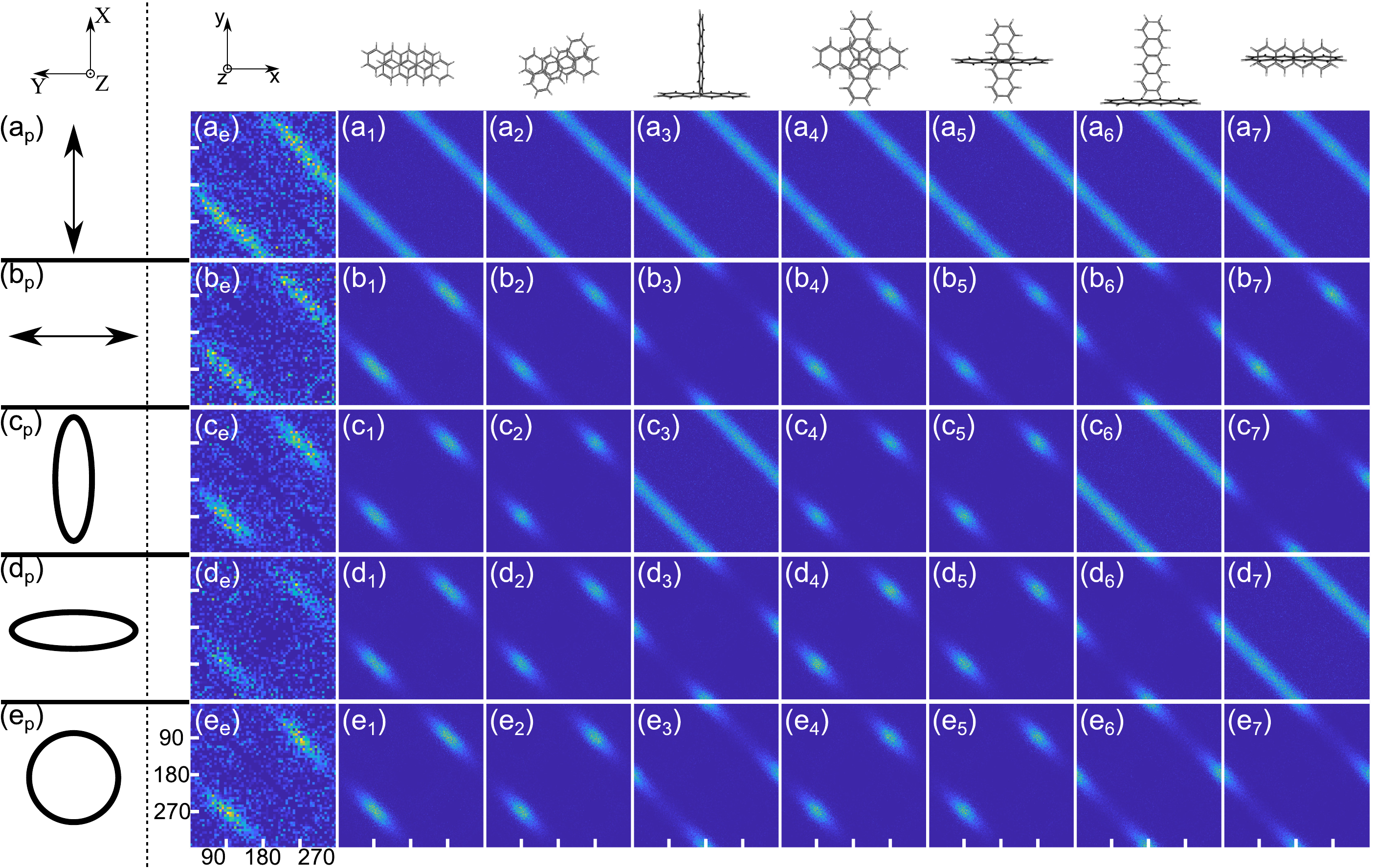}
            \caption{Covariance maps obtained for the \ce{Tc} dimer.
            Indices a, b, c, d, e refer to the
          alignment laser pulse polarization used as linear
          perpendicular, linear parallel, elliptical perpendicular,
          elliptical parallel, and circular, respectively, as shown
          in the first column with index p (the laboratory axes are illustrated at the top ).
           Perpendicular and parallel refer to the angle between the main polarization
          axis and the plane of the detector. Index e (second column from the left)
          refers to the covariance maps obtained in the experiments,
          while indices 1--7 refer to different candidate
          conformations of the tetracene dimer depicted on top of the
          corresponding column. Each panel axis ranges linearly from 0
          to 360$^\circ$.}
    \label{Fig:Covariance}
\end{figure*}

To identify possible conformations of the dimer that can produce the covariance maps observed, we simulated covariance maps for seven archetypal dimer conformations shown at the top of \autoref{Fig:Covariance}. The first two conformers are those predicted by our computational modeling to be stable in the gas phase, with the tetracene monomers parallel to each other and either parallel displaced (conformation 1) or slightly rotated
(conformation 2). The other five conformations were chosen as representative examples of other possible geometries. Although the computational modelling does not predict them to be stable in the gas phase they may get trapped in shallow local energy minima in the presence of the cold He environment as previously observed for e.g. the HCN trimers and higher oligomers.~\cite{nauta_molecular_1999,nauta_nonequilibrium_1999}

The strategy we applied to simulate the covariance map for each of the dimer conformations is the following: (1) Determine the alignment distribution, either 1D or 3D, of the dimer; (2) Calculate the laboratory-frame emission angles of the \ce{Tc^+} ions for each dimer conformation assuming Coulomb repulsion between two singly charged monomers using partial charges on each atomic center; (3) Determine the angular distribution in the detector plane, $\mathbf{X}$; (4) Calculate the covariance map, Cov($\mathbf{X}$,$\mathbf{X}$) which can be compared to experimental findings. The details of each of the four steps of the strategy is outlined in the appendix.

Starting with the alignment pulse linearly polarized along the X axis [row (a)], all proposed conformers are found to produce stripes centered at $\theta_2 = \theta_1 \pm 180{\degree}$. This is the same as in the experimental covariance map, making these covariance maps unable to distinguish between the proposed dimer structures. The second case is where the alignment pulse is linearly polarized along the Y axis [row (b)]. All conformations, except number 3 and 6, lead to covariance islands localized around (\SI{90}{\degree},\SI{270}{\degree}) and (\SI{270}{\degree},\SI{90}{\degree}) as in the experimental data. In contrast, conformations 3 and 6 lead to covariance islands localized around (\SI{0}{\degree}, \SI{180}{\degree}) and (\SI{180}{\degree},\SI{0}{\degree}). Such covariance maps are inconsistent with the experimental observations and, therefore, conformations 3 and 6 can be discounted. To understand the covariance maps resulting from conformations 3 and 6, we note that their main polarizability components are $\alpha_{yy} > \alpha_{xx} > \alpha_{zz}$) (see \autoref{table:Polarizability}). The linearly polarized alignment field will lead to alignment of the molecular y axis along the polarization axis (the Y axis). Upon Coulomb explosion, the two \ce{Tc^+} ions will thus both be ejected along the polarization axis of the alignment pulse, i.e. along \SI{0}{\degree} and \SI{180}{\degree}.



In row (c), the alignment pulse is elliptically polarized with the major (minor) polarization axis parallel to the X axis (Y axis). The covariance maps for conformations 1, 2, 4 and 5 show islands at (\SI{90}{\degree},\SI{270}{\degree}) and (\SI{270}{\degree},\SI{90}{\degree}) similar to the experimental data in \autoref{Fig:Covariance}(c$_\text{e}$). In contrast, the covariance map for conformation 7 contains two islands centered at (\SI{0}{\degree},\SI{180}{\degree}) and (\SI{180}{\degree},\SI{0}{\degree}) and therefore, we discard this conformation among the candidates for the experimental observations. The polarizability components of conformation 7 are $\alpha_{xx} > \alpha_{zz} > \alpha_{yy}$. The elliptically polarized field will align the x axis along the X axis and the z axis along the Y axis. Following Coulomb explosion the two \ce{Tc^+} ions will thus both be ejected along the minor polarization axis (the Y axis).


In row (d), the alignment pulse is elliptically polarized but now with the major (minor) polarization axis parallel to the Y axis (X axis). Again, the covariance maps for conformations 1, 2, 4 and 5 show islands at (\SI{90}{\degree}, \SI{270}{\degree}) and (\SI{270}{\degree}, \SI{90}{\degree}) similar to the experimental data in \autoref{Fig:Covariance}(d$_\text{e}$). In contrast, the covariance islands for conformations 3, 6 and 7 differ from the experimental results. Since these three conformations have already been eliminated, the covariance maps in row (d) do not narrow further the possible candidates for the dimer conformation(s).

In row (e) the alignment pulse is circularly polarized. Once again, the covariance maps for conformations 1, 2, 4 and 5 show islands at (\SI{90}{\degree},\SI{270}{\degree}) and (\SI{270}{\degree},\SI{90}{\degree}) similar to the experimental data in \autoref{Fig:Covariance}(d$_\text{e}$). Thus, the covariance maps resulting from molecules aligned with the circularly polarized pulse do not allow us to eliminate additional conformations of the tetracene dimer. At this point, we are left with conformations 1, 2, 4, and 5, which all present covariance maps consistent with the experimental data.

In the case of the \ce{OCS} and \ce{CS2} dimers, an additional experimental observable, besides the parent ions, was available for further structure determination, namely the \ce{S+} ion resulting from Coulomb fragmentation of the molecular monomers.~\cite{pickering_alignment_2018,pickering_femtosecond_2018} Fragmentation of \ce{Tc} will result in \ce{H+} or hydrocarbon fragments. Both \ce{H+} and hydrocarbon fragment ions can originate from different parts of the \ce{Tc} molecule and thus their angular distributions may be less useful for extracting further structural information of the dimer than what was the case for the \ce{S+} ions in the previous studies of \ce{OCS} and \ce{CS2}. Instead, we performed an alternative type of measurements by recording the alignment-dependent ionization yield of the \ce{Tc} dimer. As described in the next section, such measurements allow us to discount further conformations.


\subsection{Ionization anisotropy}
\label{Sec:Ionization anisotropy}

Previous works, experimentally as well as theoretically, have shown that the rate of ionization of molecules induced by intense linearly polarized laser pulses depends strongly on the alignment of the molecules with respect to the polarization direction of the pulse.~\cite{kjeldsen_influence_2005,petretti_alignment-dependent_2010,hansen_orientation-dependent_2012,mikosch_channel-_2013} In this section, we use the alignment-dependent ionization rate of the tetracene dimers to infer further information about their possible conformation. The starting point is to characterize the alignment-dependent yield of \ce{Tc+} ions produced when the tetracene monomers are ionized by the probe pulse. In practice, this involves using the monomer doping condition, similar to that used for the data presented in \autoref{fig:ion-images-covariance}, row (a), and, furthermore, analyzing only the low kinetic energy \ce{Tc+} ions stemming primarily from ionization of monomers. The measurements were performed with the alignment pulse linearly polarized either parallel or perpendicular to the probe pulse polarization and as a function of the delay between the two pulses.

\begin{figure}
  \centering
  \includegraphics[scale=1]{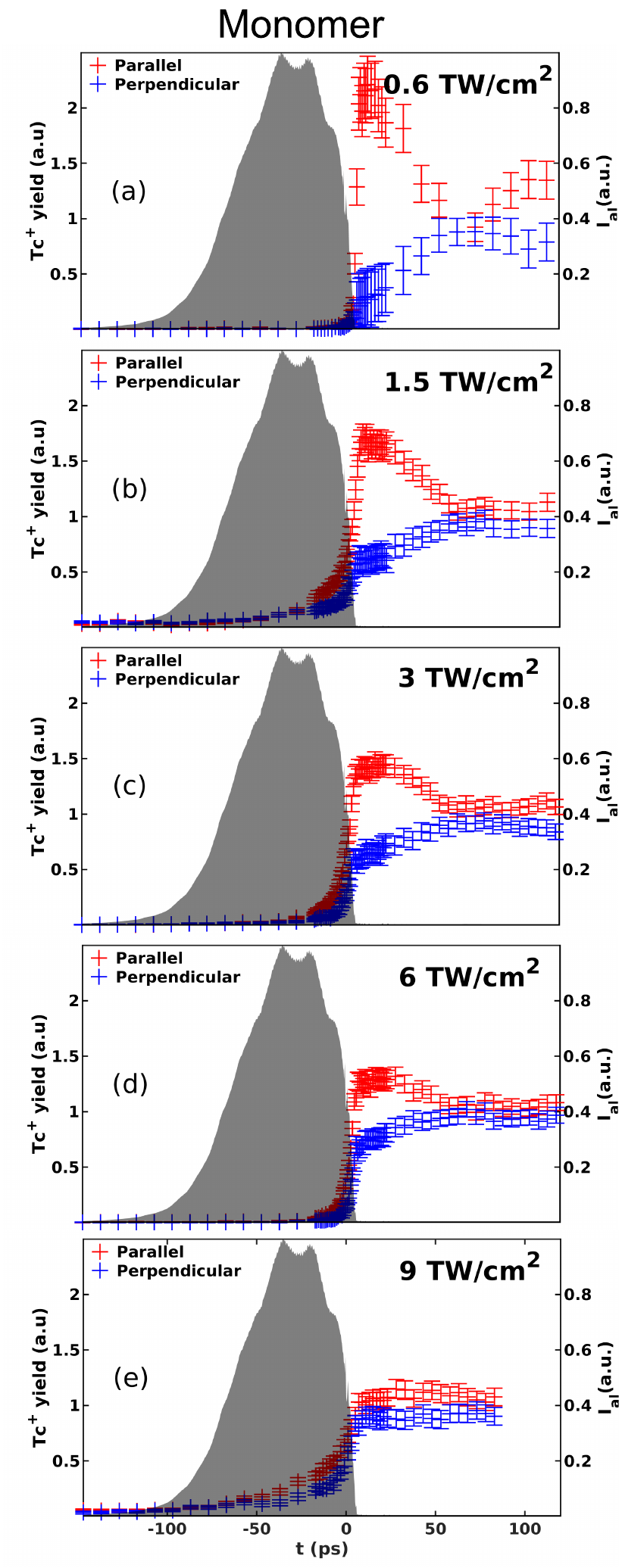}
  \caption{Time-dependent yield of \ce{Tc+} ions originating from ionization of 1D aligned tetracene molecules at different probe laser pulse
    intensities written in bold inside each panel, with the linearly polarized probe pulse parallel (red) or perpendicular (blue) to the alignment pulse polarization. In each panel the shaded area represents the intensity profile of the alignment pulse obtained by cross-correlation
    with the probe laser pulse. In each panel the \ce{Tc+} ion yield has been normalized to the mean of the yields obtained in the parallel and the perpendicular geometries at times longer than \SI{55}{ps}.
}
\label{Monomeryield}
\end{figure}

The results obtained for five different intensities of the probe pulse, $\text{I}_{\rm probe}$ are shown in \autoref{Monomeryield}. In all five panels, the \ce{Tc+} signal is very low, almost zero, when the ionization occurs while the alignment pulse is still on. The reason is that the \ce{Tc+} ions produced can resonantly absorb one or several photons from the alignment pulse, which will lead to fragmentation, i.e. destruction of intact tetracene parent ions. Previously, similar observations were reported for other molecules.~\cite{Chatterley2019,boll_imaging_2014} To study the alignment-dependent ionization yield, using \ce{Tc+} ions as a meaningful observable, it is therefore necessary to conduct measurement after the alignment field is turned off. At $t=10$~ps, the intensity of the alignment pulse is reduced to $0.5\%$ of its peak value. This is sufficiently weak to avoid the destruction of the \ce{Tc+} ions and, crucially, at this time the degree of alignment is still expected to be almost as strong as at the peak of the alignment pulse.~\cite{Chatterley2019} The red and blue data points show that for $\text{I}_{\rm probe}$ = \SI{0.6}{TW/cm^2}, the ionization yield is a factor of $\sim 7$ times higher for the parallel compared to the perpendicular geometry. In other words, the cross section for ionization is higher when the probe pulse is polarized parallel rather than perpendicular to the long axis of the tetracene molecule. This observation is consistent with experiments and calculations on strong-field ionization of related asymmetric top molecules like naphthalene, benzonitrile and anthracene.~\cite{hansen_orientation-dependent_2012,johansen_generation_2016}

At longer times, the \ce{Tc+} signal for the parallel geometry decreases , reaches a local minimum  around $t=70$~ps and then increases slightly again, while the perpendicular geometry shows a mirrored behavior. This behavior is a consequence of the time-dependent degree of alignment induced when the alignment pulse is truncated.~\cite{Chatterley2019} Finally, panels (b)-(e) of \autoref{Monomeryield} show that the contrast between the \ce{Tc+} yield in the parallel and the perpendicular geometries at $t=10$~ps decreases as $\text{I}_{\rm probe}$ is increased. We believe this results from saturation of the ionization process.


\begin{figure}
  \centering
  \includegraphics[scale=1]{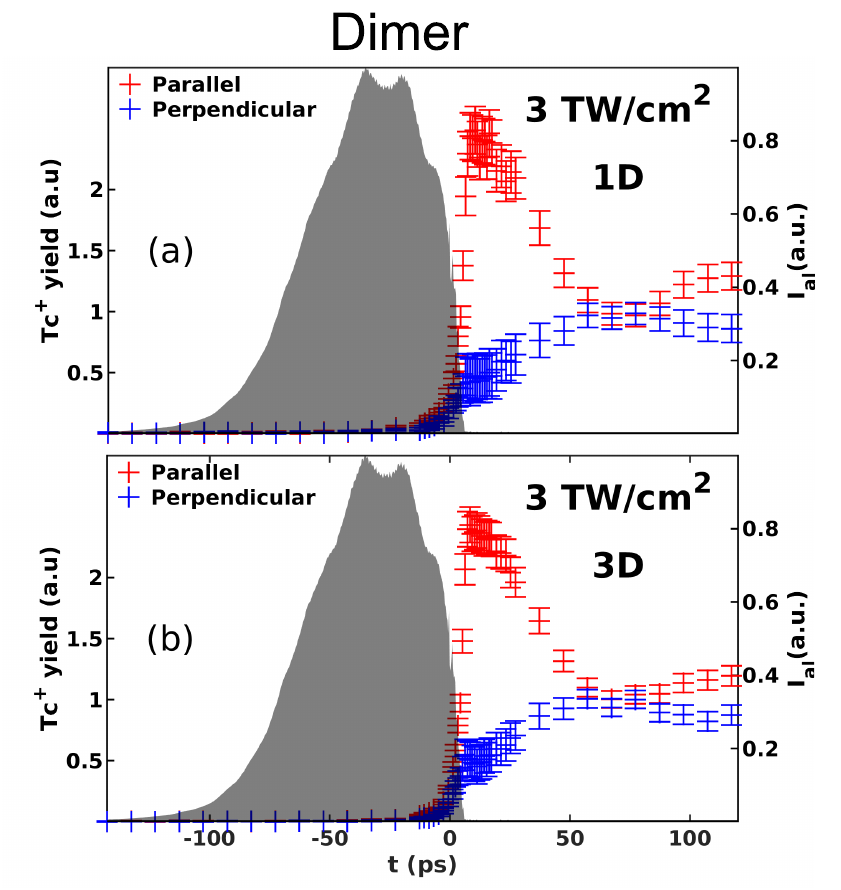}
  \caption{Time-dependent yield of \ce{Tc+} ions originating from ionization of (a) 1D aligned tetracene dimers; (b) 3D aligned tetracene dimers.  The meaning of the perpendicular and parallel curves and the shaded areas is the same as in \autoref{Monomeryield}. In both panels the \ce{Tc+} ion yield has been normalized to the mean of the yields obtained in the parallel and the perpendicular geometries at times longer than \SI{55}{ps}.}
  \label{Dimeryield}
\end{figure}

Next, similar measurements were conducted for the tetracene dimer by using the dimer doping condition, i.e. as for the data presented in \autoref{fig:ion-images-covariance} row (b), and analyzing only the high kinetic energy \ce{Tc+} ions stemming from ionization of dimers. The intensity of the probe pulse was set to \SI{3}{TW/cm^2} rather than \SI{0.6}{TW/cm^2}, in order to obtain a sufficient probability for ionizing both \ce{Tc} monomers in the dimers. Panel (a) shows the result for 1D aligned monomers. The time dependence of \ce{Tc+} ion yield is very similar to that recorded for the \ce{Tc} monomer at the same probe intensity, \autoref{Monomeryield}(c), for both the parallel and the perpendicular polarization geometries. In fact, the ratio of the \ce{Tc+} ion yield in the parallel and the perpendicular geometries at $t=10$~ps is $\sim$ 5.5, which is even larger than the ratio recorded for the monomer, $\sim$ 2.3. Such a significant difference in the ionization yield for the parallel and perpendicular geometry implies that the structure of the dimer must be anisotropic, as in the case of the monomer, i.e. possesses a 'long' axis leading to the highest ionization rate when the probe pulse is polarized along it. Such an anisotropic structure is compatible with conformations 1 and 2 but not with conformations 4 and 5. The experiment was repeated for 3D aligned molecules. The results, displayed in \autoref{Dimeryield}(b) are almost identical to those obtained for the 1D aligned molecules. They corroborate the conclusion made for the 1D aligned case but they do not allow us to distinguish between conformations 1 and 2.


\subsection{Discussion}

The comparison of the experimental covariance maps to the calculated maps leaves us with conformations 1 and 2, the two lowest-energy structures predicted by our gas-phase calculations.
We note that the rotated conformers obtained by DFT
minimization may also possess an offset, where the two centers of charge do not
lie on a common axis perpendicular to the molecular
planes. Calculations yield offsets between the charge centers ranging between 0.9 and
1.3~\AA\ depending on the method, conformers optimized with the quantum force
field being almost symmetric with values below 0.01~\AA. For
comparison, the offset in the parallel displaced conformer is closer
to 2~\AA. The rotated conformers predicted here are thus also partly
shifted. For the application of measuring singlet fission effects, this offset is essential, as perfect stacking of chromophores leads to a cancellation of the interactions singlet fission relies on.~\cite{smith_singlet_2010,smith_recent_2013} These results thus suggest that helium nanodroplets may be a fruitful route to exploring singlet fission processes.

The quantum chemistry calculations were carried out for isolated tetracene dimers. Although the interaction with helium was expected to be
negligible in rationalizing the conformations of the tetracene dimer, it may still influence the dynamical formation when the two tetracene molecules are picked up in the droplet. The helium solvent is known to be attracted more strongly to the hydrocarbon and somewhat freeze at its contact,~\cite{heidenreich01} possibly resulting in snowball precursors.~\cite{calvo12} Such effects may even be magnified in the presence of multiple molecules~\cite{calvoyb16} and it is possible that commensurate conformers such as the parallel
displaced structure mostly identified in our experiment may be
kinetically favored once embedded in helium droplets.

As discussed in \autoref{sec:comparison-ang-cov-maps}, comparison of the experimental angular covariance maps for \ce{Tc+} to the calculated maps do not allow us to distinguish between the rotated-parallel and slipped-parallel conformations --- and this would also be the case for the slipped-and-rotated-parallel conformation. However, if an atom, like \ce{F}, was substituted for one of the hydrogens in each \ce{Tc} molecule, then distinction between the conformations might become feasible by observing the relative angle between the emission of \ce{F+} ions from the two monomers. Previous experiments on halogen-atom substituted biphenyls in the gas phase have shown that angular covariance maps, generated from recoiling halogen ions following Coulomb explosion, are well suited for determination of bond angles and dihedral angles.~\cite{slater_covariance_2014,christensen_dynamic_2014} We believe transfer of this methodology to dimers of halogenated PAHs embedded in He droplets is feasible and promising for structure determination, including time-resolved measurements.

\section{Conclusions}

The purpose of this study was to obtain information about the conformation of tetracene dimers in a combined experimental and theoretical study. Experimentally, tetracene dimers were formed inside He nanodroplets. A strong femtosecond probe laser pulse was used to ionize both \ce{Tc} molecules in the dimer, leading to a pair of recoiling \ce{Tc+} cations resulting from their internal Coulomb repulsion. These kinetic \ce{Tc+} ions provided an experimental observable uniquely sensitive to droplets doped with dimers. Next, a slow turn-on, fast turn-off, moderately intense laser was used to create a window of field-free alignment shortly after the pulse. Synchronizing the probe pulse to this window, the dimers, aligned either 1-dimensionally or 3-dimensionally, were Coulomb exploded and the covariance map of the emission directions of the \ce{Tc+} recoil ions determined. As a reference, angular covariance maps were calculated for seven different conformations including the two predicted to be the most stable by our quantum chemistry calculations and another five chosen as representative examples of other possible geometries. The experimental angular covariance maps were found to be consistent with four of the calculated maps. An additional dimer structure sensitive measurement was conducted, namely how the yield of strong-field ionization depends on the polarization axis of the probe pulse with respect to the alignment of the dimer. It was found that the ionization yield is a factor of five times higher when the probe pulse polarization was parallel compared to perpendicular to the most polarizable axis of the dimer. This result is only consistent with the two tetracene molecules in the dimer being parallel to each other and either slightly displaced or slightly rotated. These are the two most stable gas-phase conformations of the dimer according to our quantum chemistry calculations.


\begin{acknowledgments}
We acknowledge support from the following three funding sources: The European Union's Horizon 2020 research and innovation programme under the Marie Sklodowska-Curie grant agreement No 674960. A Villum Experiment Grant (no. 23177) from The Villum Foundation. The European Research Council-AdG (Project No. 320459, DropletControl). We thank Frank Jensen for theoretical support.
\end{acknowledgments}

\appendix

\section*{Appendices}
\label{section:appendix}

\subsection*{Simulation procedure}

The identification of dimer conformations from the covariance maps
relies on comparison with simulated maps predicted for different
candidate structures. The individual tetracene
monomers are expected to be rigid and the problem amounts to
finding the relative orientations between the two molecules and the
possible shift between their centers of mass.

For each conformation considered, covariance maps were generated by
extracting the projection of the separation distance ($\vert\vec{r}_{\rm diff}\vert$) between the center of mass of each tetracene molecule on
the detector plane for different polarizations of the alignment laser
pulse. The separation distance indicates the direction of recoil that
the two \ce{Tc+} should follow by repelling each other. This section
details the overall procedure.

\subsection*{Conformations}
\label{ssection:Conformations}
One important ingredient in explaining the observed covariance maps is
the polarizability tensor of the tetracene dimer, which is responsible
for its interaction with the laser pulses and is determined by its
structure.

The polarizability of the tetracene monomer was calculated with a density
functional theory method (wB97xD)~\cite{B810189B} using the diffuse
basis set aug\hyp{}pcseg\hyp{}n,~\cite{doi:10.1002/anie.200300611} after a geometry optimization with a similar method but a more
localized basis set pcseg\hyp{}n.~\cite{doi:10.1021/ct401026a} These
calculations performed with the Gaussian09 software package~\cite{G09}
yielded the polarizability components as
$\alpha_{xx}=$\SI{63.1}{\angstrom^3},
$\alpha_{yy}=$\SI{31.9}{\angstrom^3}, $\alpha_{zz} =
$\SI{15.9}{\angstrom^3}, where $x$, $y$, $z$ refer to the major, minor and
orthogonal axis of tetracene, respectively.

The polarizability tensor of the dimer is sometimes assumed to be the sum of the
polarizability tensor of each molecule. However, this approximation usually
underestimates the interaction polarizability along the axis
connecting the molecules, while overestimating polarizability
perpendicular to it.~\cite{doi:10.1063/1.1433747} Simulations carried
out for urea (\ce{CH4N2O}) and fullerenes reported an increase of the
polarizability component along the dimer axes by 9.9\% and 17.8\%,
respectively, while the other components displayed a decrease by a few
percents only.~\cite{doi:10.1063/1.1433747} In our case, the anisotropy in
the polarizability tensor of tetracene could be considered as sufficiently large that we can
neglect these effects and assume the polarizability of the dimer to be a
linear sum, although some caution should be taken with some of the
conformations presented below. To ascertain this ambiguity, we calculated the exact polarizability for each conformation shown in \autoref{table:Polarizability} using a fixed geometry and a similar method and basis as the one mentioned for the monomer. We see that using the exact polarizabilities breaks the symmetry that some of the dimers would posses if only the sum of their polarizabilities were considered. We thus only use the exact polarizabilities for the simulations presented above.
\begin{table}
\centering
\begin{tabular}{|c|c|c|}
\hline
Conformation & \multicolumn{2}{c|}{(Polarizability tensors}\\
                         \cline{2-3}
							& Approximate & Exact \\
							\hline
1 & $\begin{pmatrix} 126.2 & 0 & 0\\ 0 & 47.9&0\\0&0& 47.9 \end{pmatrix}$ & $\begin{pmatrix} 104.0 & -24.9 & 1.0 \\-24.9 & 64.5 & 1.5 \\ 1.0 & 1.5 & 44.1 \end{pmatrix}$\\
\hline
2 & $\begin{pmatrix} 126.2 & 0 & 0\\ 0 & 47.9&0\\0&0& 47.9 \end{pmatrix}$ & $\begin{pmatrix} 91.5 & -12.0 & 0.5\\ -12.0 & 78.6 & 0  \\0.5& 0 & 34.9\end{pmatrix}$\\
\hline
3 & $\begin{pmatrix}79.0 & 0 & 0\\ 0 & 79.0&0\\0&0& 63.8 \end{pmatrix}$ & $\begin{pmatrix} 74.5 & 0 & 0\\ 0 & 97.3&0\\0&0& 57.1\end{pmatrix}$\\
\hline
4 & $\begin{pmatrix} 95.0 & 0 & 0\\ 0 & 95.0&0\\0&0& 31.9 \end{pmatrix}$ & $\begin{pmatrix} 84.3 & 0 & 0\\ 0 & 84.3&0\\0&0& 34.4\end{pmatrix}$\\
\hline
5 & $\begin{pmatrix} 95.0 & 0 & 0\\ 0 & 79.0&0\\0&0& 47.9\\\end{pmatrix}$ & $\begin{pmatrix} 85.4 & 0 & 0\\ 0 & 74.0&0\\0&0& 57.4\end{pmatrix}$\\
\hline
6 & $\begin{pmatrix} 95.0 & 0 & 0\\ 0 & 79.0&0\\0&0& 47.9\\  \end{pmatrix}$ & $\begin{pmatrix} 85.6 & 0 & 0\\ 0 & 94.1&0\\0&0& 43.2 \end{pmatrix}$\\
\hline
7 & $\begin{pmatrix} 126.2 & 0 & 0\\ 0 & 47.9&0\\0&0& 47.9 \end{pmatrix}$ & $\begin{pmatrix} 104.0 & 0 & 0\\ 0 & 41.6&0\\0&0& 60.9\end{pmatrix}$\\
\hline
\end{tabular}
\caption{\label{table:Polarizability} Table listing polarizability tensors for each conformer presented in \autoref{Fig:Covariance}. The polarizability components are expressed in $\AA^3$ and are listed in the (x,y,z) order. The "Approximate" column comes from the summation of the polarizability tensor of each monomer. The "Exact" column contains values resulting from DFT calculations. The results presented in \autoref{Fig:Covariance} used the exact values.}
\end{table}

\subsection*{Laser-induced alignment}

After the diagonalization of the polarizability tensors obtained for each conformation, it is possible to calculate
the potential energy surface when an electric field is applied. The potential can be evaluated using second-order perturbation theory~\cite{Atkins} and is defined as:
\begin{equation}
V = -\dfrac{1}{2} \vec{E}^\mathsf{T}\bm\alpha_{\rm dimer}\vec{E}= -\dfrac{1}{2} \vec{\mu}_{ind}^\mathsf{T}\cdot\vec{E},
 \label{Potential}
\end{equation}
where the superscript $\sf{T}$ stands for transpose, $\bm{\alpha}_{\rm dimer}$ is the polarizability tensor of the conformer, $\vec{\mu}_{ind}$ is the
induced dipole moment and $\vec{E}$ the electric field applied on the system.

In our case, the electric field comes from a strong non-resonant laser pulse, and can be expressed as:
\begin{equation}
\vec{E}_{\rm laser}=
\dfrac{E_0}{\sqrt{1+\epsilon^2}} \begin{pmatrix}\cos\omega t \\
\pm \epsilon \sin\omega t\\
0
\end{pmatrix}.
\label{EField}
\end{equation}
In this equation, $E_0$ is the peak amplitude of the electric field that we can extract from the experiment by measuring the intensity of the laser pulse, $\epsilon$ is the ellipticity
parameter which ranges from 0 (linear polarization) to 1 (circular polarization), and $\omega$ is the laser frequency.

Usually, the Schrödinger equation is then solved and the distribution of rotational states that will be populated in the presence of the electric field is calculated to extract the overall angular distribution of the complex. However, in our approach we drastically simplify the problem by making use of the adiabaticity of the alignment process, which will result in an angular confinement of the complex at the bottom of the potential well presented in Eq.\eqref{Potential}. The non-perfect alignment is accounted for by addition of a spread in the angular distribution.

To connect the $\vec{E}_{laser}$ in Eq.\eqref{EField} to the electric field $\vec{E}$ used in Eq.\eqref{Potential}, we need both $\vec{E}_{laser}$ and $\bm\alpha_{\rm dimer}$ to be expressed in the same frame.
In our case, we decided to make use of the diagonal form of the polarizability tensor in the molecular frame (MF) and to calculate the potential surface in this frame. For this purpose, we used a uniform distribution for all the possible directions of the $\vec{E}_{\rm laser}$ in the MF. A uniform sphere made of 10\,000 directions vectors was created to represent the main axis of the polarization ellipse. This would be sufficient for a linearly polarized pulse but not for an elliptically polarized pulse, where the minor axis has to be included. In this case, for each major axis orientation 1000 angular steps of the minor axis were used.

To make the procedure explicit and to give a more intuitive form for the interaction potential, we develop the expression using a general form for $\vec{E}_{\rm laser}$ in the molecular frame:
\begin{equation}
\vec{E}_{\rm laser}=
\dfrac{E_0}{\sqrt{1+\epsilon^2}} \begin{pmatrix}
a_x\cos\omega t +b_x \epsilon \sin\omega t\\
a_y\cos\omega t +b_y \epsilon \sin\omega t\\
a_z\cos\omega t +b_z \epsilon \sin\omega t
\end{pmatrix}.
\label{EFieldMF}
\end{equation}
where $a_i$ and $b_i$ are coefficients fulfilling the conditions:
\begin{eqnarray}
\sum a_{i}^2 = \sum b_{i}^2 = 1 \label{condition}
\\
\sum a_ib_i = 0
\end{eqnarray}
These coefficients will depend on the unitary transformations that have been applied to $\vec{E}_{\rm laser}$ and will be the sum of trigonometric functions with the angles related to the applied rotations.
Implementing this expression in Eq.\eqref{Potential} and developing it using the long duration of the pulse compared to the optical frequency, we obtain:
\begin{eqnarray}
\langle V\rangle_{T}  = - \dfrac{E_0^2}{4\left(1+\epsilon^2\right)}\sum_{i = x,y,z}\left(a_i^2 + b_i^2\epsilon^2\right)\alpha_{ii}
\end{eqnarray}
This shape is interesting since only square quantities appear in the summation.
To develop the expression a bit further, some assumptions are needed about the shape of the polarizability tensor. Assuming that two of its components are larger than the last one ($\alpha_{xx},\alpha_{yy} \gg \alpha_{zz}$), the minimum of the potential will be found if one tries to maximize the coefficient $a_{x,y}$ and $b_{x,y}$ which will naturally lead to put $a_z = b_z = 0$ thanks to condition of Eq.\eqref{condition}.
A consistent form with the previous requirements would then be:
\begin{eqnarray}
\begin{pmatrix}
a_x\\
a_y\\
a_z
\end{pmatrix} =
\begin{pmatrix}
\cos\phi\\
-\sin\phi\\
0
\end{pmatrix}
\: , \:
\begin{pmatrix}
b_x\\
b_y\\
b_z
\end{pmatrix} =
\begin{pmatrix}
\sin\phi\\
\cos\phi\\
0
\end{pmatrix}
\end{eqnarray}
thus giving

\begin{eqnarray*}
\langle V\rangle_{T} &=&
-\dfrac{E_0^2}{8}\left[\left(\alpha_{xx}+\alpha_{yy}\right) +
  \left(\alpha_{xx}
  -\alpha_{yy}\right)\dfrac{1-\epsilon^2}{1+\epsilon^2}\cos
  2\phi\right].
\end{eqnarray*}
In the extreme case of $\epsilon =0$ (linear pulse), the minimum of
$V$ is achieved when $\phi =0 [\pi]$ if $\alpha_{xx}>\alpha_{yy}$ or
$\phi =\pi/2 [\pi]$ if $\alpha_{yy}>\alpha_{xx}$. With $\epsilon =
1$ (circular pulse) or if $\alpha_{xx} = \alpha_{yy}$, the dependence in $\phi$ disappears, leaving an
isotropic distribution in the plane (all $\phi$ are allowed).

\subsection*{Frames connection}

Picking the orientation of the electric field that gives the lowest
potential energy, the minima are chosen using a spread ($\Delta E_{\rm lim}$)
based on the temperature of the dimers inside the helium droplets
(\SI{0.37}{K}). The energy spread  $\Delta E_{\rm lim}$ is chosen to include $99\%$
of the Boltzmann populations, which gives $\Delta E_{\rm lim} =
\SI{0.15}{meV}$. Only energies fulfilling $E <E_{\rm min} + \Delta E_{\rm lim}$ are selected, with $E_{\rm min}$ being the lowest possible value in the
potential energy surface. \\

For each orientation found above, a rotation matrix connecting the initial frame (IF) to the laboratory frame (LF) can be defined, labeled $\bm R_{\rm IFtoLF}$.
This matrix is generated by combining two rotation matrices, the first one connects the IF where the dimer has been described to the MF where its polarizability tensor is diagonal, the second one is generated by finding the rotation needed to project the orientation of the electric field (both major and minor axes) in the LF where it should be fixed as in the experiment.
These two rotations applied on the IF allow us to express the inter-dimer distance in the LF.


\subsection*{Projection}
\label{ssection:Projection}

The direction of the repulsion vector is calculated from the resulting Coulombic force:
\begin{equation}
\vec{F}_C = \sum_i\sum_j q_iq_j\dfrac{\vec{r}_{i}-\vec{r}_{j}}{\vert\vec{r}_{i}-\vec{r}_{j}\vert^3}
\end{equation}
where $q_i$ and $q_j$ are the partial charges on each atom (shown in \autoref{Fig:tetracenemonomer}), $\vec{r}_i$ and $\vec{r}_j$ are their respective positions and the summation is carried out to consider all pairs combination between the two molecules.
Its direction in the laboratory frame is given from the
vector $\vec{F}^{\rm LF}_{\rm C} = \bm R_{\rm IFtoLF}\vec{F}^{\rm
  IF}_{\rm Coulomb}$. Since multiple solutions are
possible, a random selection over 100\,000 molecules, using a Boltzmann weight, was used to choose which $\bm R_{\rm IFtoLF}$
to apply to $\vec{F}^{\rm IF}_{\rm C}$. To take into account
imperfect alignment, the direction vector of $\vec{F}^{\rm LF}_{\rm
  C}$ was then rotated using a conical distribution described by a
polar angle $\psi$ picked from a Gaussian distribution with standard
deviation $\sigma_{\psi} =\pi/6$. This corresponds to an alignment
distribution of $\vec{F}^{\rm IF}_{\rm C}$ characterized by $\langle\cos^2\psi\rangle = 0.8$.

The vector $\vec{F}_{\rm C}$ is then projected onto the detector plane leading to two solutions, referring to the detector plane being parallel to the main axis of the polarization or perpendicular to it. This is represented in \autoref{Fig:ExpSetup} with the
main polarization axis of the alignment pulse being similar to the Y axis and the
minor axis to the X axis, or the main axis to the X axis and the minor
to the Y axis, respectively. For each solution, an angle $\theta_{\rm 2D}$ is extracted by projecting it onto the polarization axis
parallel to the detector plane. A maximal velocity can be estimated from the
magnitude of the separation vector ($\vert \vec F_{\rm C}\vert$ based on the resulting Coulomb repulsion between the partial charges:
\begin{equation}
V_{\rm Coulomb} = \sum_i\sum_j\dfrac{q_iq_j}{2\vert \vec{r}_{i}-\vec{r}_{j}\vert} = K =
\dfrac{mv_{\rm max}^2}{2} \rightarrow \vert \vec v_{\rm max}\vert =
\sqrt{\dfrac{2}{m}V_{\rm Coulomb}},
\end{equation}
with $K$ the kinetic energy, $m$ the mass of the molecule and $v_{\rm max}$ is the maximum
velocity that the system can acquire upon Coulomb
repulsion. The Coulomb potential has been divided by two to take into account the symmetric behavior of the two molecules during Coulomb explosion.
  The initial distance between the molecules will depend on the conformations considered. A minimal distance of \SI{3.5}{\angstrom} was assumed with the centers of mass based on the molecular modeling in \autoref{sec:qchem} except for geometries 3 and 6 where a larger separation had to be added ($\sim$\SI{6}{\angstrom}) to avoid unphysical overlap.
Owing to the screening of low kinetic energy ions during the
experiment (black circle area in \autoref{fig:ion-images-covariance}), a filter can be
applied on the velocity $v$, and values below $v_{\rm lim}$ are
excluded. In our case, $v_{\rm lim} = $\SI{0.25}{mm/\mu s} as estimated from simulations with SIMION~\cite{lapack99} giving the expected velocities as a function of the spectrometer radius.

\subsection*{Covariance}

For a given dimer conformation, the covariance map is eventually
generated using the angle $\theta_{\rm 2D}$ coming from the
projection. This angle is selected with a uniform distribution
weighted by its number of occurrences, as explained in the previous
section. For each selected angle $\theta^i_{\rm 2D}$, a mirror angle
$\theta^j_{\rm 2D} = \theta^i_{\rm 2D} + 180^\circ$ is generated as
well to account for the recording of two molecules emitted upon
Coulomb dissociation of a given dimer and producing two counts. A
Gaussian noise with standard deviation $\sigma_{\theta_{\rm 2D}}$ was
also applied on each angle to take into account the non-axial
recoil.~\cite{doi:10.1063/1.1147588} Here $\sigma_{\theta_{2D}}$ was
extracted from experimental data by measuring the width in the
covariance stripes originating from unaligned dimers.~\cite{Lauge}
Using this procedure, we found $\sigma_{\theta_{2D}} \approx
\SI{15}{\degree}$.

%




\begin{thebibliography}{10}
\newcommand{\enquote}[1]{``#1''}

\bibitem{hobza_world_2006}
P.~Hobza, R.~Zahradn{\'i}k, and K.~M{\"u}ller-Dethlefs.
\newblock \enquote{The {World} of {Non}-{Covalent} {Interactions}: 2006}.
\newblock Collect. Czech. Chem. Commun., \textbf{71(4)}, 443--531 (2006).

\bibitem{a.ikkanda_exploiting_2016}
B.~A.~Ikkanda and B.~L.~Iverson.
\newblock \enquote{Exploiting the interactions of aromatic units for folding
  and assembly in aqueous environments}.
\newblock Chem. Commun., \textbf{52(50)}, 7752--7759 (2016).

\bibitem{becucci_high-resolution_2016}
M.~Becucci and S.~Melandri.
\newblock \enquote{High-{Resolution} {Spectroscopic} {Studies} of {Complexes}
  {Formed} by {Medium}-{Size} {Organic} {Molecules}}.
\newblock Chem. Rev., \textbf{116(9)}, 5014--5037 (2016).

\bibitem{smith_singlet_2010}
M.~B. Smith and J.~Michl.
\newblock \enquote{Singlet {Fission}}.
\newblock Chemical Reviews, \textbf{110(11)}, 6891--6936 (2010).

\bibitem{nesbitt_high-resolution_1988}
D.~J. Nesbitt.
\newblock \enquote{High-resolution infrared spectroscopy of weakly bound
  molecular complexes}.
\newblock Chem. Rev., \textbf{88(6)}, 843--870 (1988).

\bibitem{moazzen-ahmadi_spectroscopy_2013}
N.~Moazzen-Ahmadi and A.~R.~W. McKellar.
\newblock \enquote{Spectroscopy of dimers, trimers and larger clusters of
  linear molecules}.
\newblock International Reviews in Physical Chemistry, \textbf{32(4)}, 611--650
  (2013).

\bibitem{felker_rotational_1992}
P.~M. Felker.
\newblock \enquote{Rotational coherence spectroscopy: studies of the geometries
  of large gas-phase species by picosecond time-domain methods}.
\newblock J. Phys. Chem., \textbf{96(20)}, 7844--7857 (1992).

\bibitem{riehn_high-resolution_2002}
C.~Riehn.
\newblock \enquote{High-resolution pump{\textendash}probe rotational coherence
  spectroscopy {\textendash} rotational constants and structure of ground and
  electronically excited states of large molecular systems}.
\newblock Chemical Physics, \textbf{283(1{\textendash}2)}, 297--329 (2002).

\bibitem{doi:10.1146/annurev.physchem.49.1.481}
D.~W. Pratt.
\newblock \enquote{High resolution spectroscopy in the gas phase: Even large
  molecules have well-defined shapes}.
\newblock Annu. Rev. Phys. Chem., \textbf{49(1)}, 481--530 (1998).

\bibitem{doi:10.1063/1.465035}
E.~Arunan and H.~S. Gutowsky.
\newblock \enquote{The rotational spectrum, structure and dynamics of a benzene
  dimer}.
\newblock J. Chem. Phys., \textbf{98(5)}, 4294--4296 (1993).

\bibitem{doi:10.1063/1.466382}
P.~G. Smith, S.~Gnanakaran, A.~J. Kaziska, A.~L. Motyka, S.~M. Hong, R.~M.
  Hochstrasser, and M.~R. Topp.
\newblock \enquote{Electronic coupling and conformational barrier crossing of
  9,9’‐bifluorenyl studied in a supersonic jet}.
\newblock J. Chem. Phys., \textbf{100(5)}, 3384--3393 (1994).

\bibitem{doi:10.1063/1.452252}
T.~Kobayashi, K.~Honma, O.~Kajimoto, and S.~Tsuchiya.
\newblock \enquote{Benzonitrile and its van der \uppercase{W}aals complexes
  studied in a free jet. \uppercase{I}. the \uppercase{LIF} spectra and the
  structure}.
\newblock J. Chem. Phys., \textbf{86(3)}, 1111--1117 (1987).

\bibitem{SCHMITT2006234}
M.~Schmitt, M.~Böhm, C.~Ratzer, S.~Siegert, M.~van Beek, and W.~L. Meerts.
\newblock \enquote{Electronic excitation in the benzonitrile dimer: The
  intermolecular structure in the \uppercase{S}0 and \uppercase{S}1 state
  determined by rotationally resolved electronic spectroscopy}.
\newblock J. Mol. Struct., \textbf{795(1)}, 234 -- 241 (2006).

\bibitem{doi:10.1002/cphc.200500670}
M.~Schmitt, M.~Böhm, C.~Ratzer, D.~Krügler, K.~Kleinermanns, I.~Kalkman,
  G.~Berden, and W.~L. Meerts.
\newblock \enquote{Determining the intermolecular structure in the
  \uppercase{S}0 and \uppercase{S}1 states of the phenol dimer by rotationally
  resolved electronic spectroscopy}.
\newblock ChemPhysChem, \textbf{7(6)}, 1241--1249 (2006).

\bibitem{doi:10.1021/jp903236z}
G.~Pietraperzia, M.~Pasquini, N.~Schiccheri, G.~Piani, M.~Becucci,
  E.~Castellucci, M.~Biczysko, J.~Bloino, and V.~Barone.
\newblock \enquote{The gas phase anisole dimer: A combined high-resolution
  spectroscopy and computational study of a stacked molecular system}.
\newblock J. Phys. Chem. A, \textbf{113(52)}, 14343--14351 (2009).

\bibitem{choi_infrared_2006}
M.~Y. Choi, G.~E. Douberly, T.~M. Falconer, W.~K. Lewis, C.~M. Lindsay, J.~M.
  Merritt, P.~L. Stiles, and R.~E. Miller.
\newblock \enquote{Infrared spectroscopy of helium nanodroplets: novel methods
  for physics and chemistry}.
\newblock Int. Rev. Phys. Chem., \textbf{25(1)}, 15 (2006).

\bibitem{yang_helium_2012}
S.~Yang and A.~M. Ellis.
\newblock \enquote{Helium droplets: a chemistry perspective}.
\newblock Chem. Soc. Rev., \textbf{42(2)}, 472--484 (2012).

\bibitem{jaksch_electron_2009}
S.~Jaksch, I.~M{\"a}hr, S.~Denifl, A.~Bacher, O.~Echt, T.~D. M{\"a}rk, and
  P.~Scheier.
\newblock \enquote{Electron attachment to doped helium droplets:
  \uppercase{C}$_{60}$-, (\uppercase{C}$_{60}$)$_2$-, and
  (\uppercase{C}$_{60}$\uppercase{D}$_2$\uppercase{O})- anions}.
\newblock Eur. Phys. J. D, \textbf{52(1)}, 91--94 (2009).

\bibitem{wewer_molecular_2003}
M.~Wewer and F.~Stienkemeier.
\newblock \enquote{Molecular versus excitonic transitions in {PTCDA} dimers and
  oligomers studied by helium nanodroplet isolation spectroscopy}.
\newblock Phys. Rev. B, \textbf{67(12)}, 125201 (2003).

\bibitem{roden_vibronic_2011}
J.~Roden, A.~Eisfeld, M.~Dvo{\v r}{\'a}k, O.~B{\"u}nermann, and
  F.~Stienkemeier.
\newblock \enquote{Vibronic line shapes of {PTCDA} oligomers in helium
  nanodroplets}.
\newblock J. Chem. Phys., \textbf{134(5)}, 054907 (2011).

\bibitem{birer_dimer_2015}
{\"O}.~Birer and E.~Yurtsever.
\newblock \enquote{Dimer formation of perylene: {An} ultracold spectroscopic
  and computational study}.
\newblock J. Mol. Struct., \textbf{1097}, 29--36 (2015).

\bibitem{nauta_nonequilibrium_1999}
K.~Nauta and R.~E. Miller.
\newblock \enquote{Nonequilibrium {Self}-{Assembly} of {Long} {Chains} of
  {Polar} {Molecules} in {Superfluid} {Helium}}.
\newblock Science, \textbf{283(5409)}, 1895 --1897 (1999).

\bibitem{sulaiman_infrared_2017}
M.~I. Sulaiman, S.~Yang, and A.~M. Ellis.
\newblock \enquote{Infrared spectroscopy of methanol and methanol/water
  clusters in helium nanodroplets: The {OH} stretching region}.
\newblock J. Phys. Chem. A, \textbf{121}, 771--777 (2017).

\bibitem{verma_infrared_2019}
D.~Verma, R.~M.~P. Tanyag, S.~M.~O. O{\textquoteright}Connell, and A.~F.
  Vilesov.
\newblock \enquote{Infrared spectroscopy in superfluid helium droplets}.
\newblock Adv. Phys. X, \textbf{4(1)}, 1553569 (2019).

\bibitem{pickering_alignment_2018}
J.~D. Pickering, B.~Shepperson, B.~A.~K. H{\"u}bschmann, F.~Thorning, and
  H.~Stapelfeldt.
\newblock \enquote{Alignment and {Imaging} of the $\ce{CS2}$ {Dimer} {Inside}
  {Helium} {Nanodroplets}}.
\newblock Phys. Rev. Lett., \textbf{120(11)}, 113202 (2018).

\bibitem{pickering_femtosecond_2018}
J.~D. Pickering, B.~Shepperson, L.~Christiansen, and H.~Stapelfeldt.
\newblock \enquote{Femtosecond laser induced {Coulomb} explosion imaging of
  aligned {OCS} oligomers inside helium nanodroplets}.
\newblock J. Chem. Phys., \textbf{149(15)}, 154306 (2018).

\bibitem{hansen_control_2012}
J.~L. Hansen, J.~H. Nielsen, C.~B. Madsen, A.~T. Lindhardt, M.~P. Johansson,
  T.~Skrydstrup, L.~B. Madsen, and H.~Stapelfeldt.
\newblock \enquote{Control and femtosecond time-resolved imaging of torsion in
  a chiral molecule}.
\newblock J. Chem. Phys., \textbf{136(20)}, 204310--204310--10 (2012).

\bibitem{slater_covariance_2014}
C.~S. Slater, S.~Blake, M.~Brouard, A.~Lauer, C.~Vallance, J.~J. John,
  R.~Turchetta, A.~Nomerotski, L.~Christensen, J.~H. Nielsen, M.~P. Johansson,
  and H.~Stapelfeldt.
\newblock \enquote{Covariance imaging experiments using a pixel-imaging
  mass-spectrometry camera}.
\newblock Phys. Rev. A, \textbf{89(1)}, 011401 (2014).

\bibitem{frasinski_covariance_2016}
L.~J. Frasinski.
\newblock \enquote{Covariance mapping techniques}.
\newblock J. Phys. B: At. Mol. Opt. Phys., \textbf{49(15)}, 152004 (2016).

\bibitem{stapelfeldt_colloquium:_2003}
H.~Stapelfeldt and T.~Seideman.
\newblock \enquote{Colloquium: {Aligning} molecules with strong laser pulses}.
\newblock Rev. Mod. Phys., \textbf{75(2)}, 543 (2003).

\bibitem{fleischer_molecular_2012}
S.~Fleischer, Y.~Khodorkovsky, E.~Gershnabel, Y.~Prior, and I.~S. Averbukh.
\newblock \enquote{Molecular {Alignment} {Induced} by {Ultrashort} {Laser}
  {Pulses} and {Its} {Impact} on {Molecular} {Motion}}.
\newblock Isr. J. Chem., \textbf{52(5)}, 414--437 (2012).

\bibitem{doi:10.1146/annurev.astro.46.060407.145211}
A.~Tielens.
\newblock \enquote{Interstellar polycyclic aromatic hydrocarbon molecules}.
\newblock Annu. Rev. Astron. Astrophys., \textbf{46(1)}, 289--337 (2008).

\bibitem{sabbah10}
H.~Sabbah, L.~Biennier, S.~J. Klippenstein, I.~R. Sims, and B.~R. Rowe.
\newblock \enquote{Exploring the role of \uppercase{PAH}s in the formation of
  soot: Pyrene dimerization}.
\newblock J. Phys. Chem. Lett., \textbf{1}, 2962--2967 (2010).

\bibitem{smith_recent_2013}
M.~B. Smith and J.~Michl.
\newblock \enquote{Recent {Advances} in {Singlet} {Fission}}.
\newblock Annual Review of Physical Chemistry, \textbf{64(1)}, 361--386 (2013).

\bibitem{Zirzlmeier5325}
J.~Zirzlmeier, D.~Lehnherr, P.~B. Coto, E.~T. Chernick, R.~Casillas, B.~S.
  Basel, M.~Thoss, R.~R. Tykwinski, and D.~M. Guldi.
\newblock \enquote{Singlet fission in pentacene dimers}.
\newblock Proc. Natl. Acad. Sci. U. S. A., \textbf{112(17)}, 5325--5330 (2015).

\bibitem{doi:10.1021/acs.jpca.6b00988}
T.~Sakuma, H.~Sakai, Y.~Araki, T.~Mori, T.~Wada, N.~V. Tkachenko, and
  T.~Hasobe.
\newblock \enquote{Long-lived triplet excited states of bent-shaped pentacene
  dimers by intramolecular singlet fission}.
\newblock J. Phys. Chem. A, \textbf{120(11)}, 1867--1875 (2016).

\bibitem{doi:10.1063/1.4983703}
B.~Shepperson, A.~S. Chatterley, A.~A. Sondergaard, L.~Christiansen,
  M.~Lemeshko, and H.~Stapelfeldt.
\newblock \enquote{Strongly aligned molecules inside helium droplets in the
  near-adiabatic regime}.
\newblock J. Chem. Phys., \textbf{147(1)}, 013946 (2017).

\bibitem{doi:10.1002/anie.200300611}
J.~P. Toennies and A.~F. Vilesov.
\newblock \enquote{Superfluid helium droplets: A uniquely cold nanomatrix for
  molecules and molecular complexes}.
\newblock Angew. Chem. Int. Ed., \textbf{43(20)}, 2622--2648 (2004).

\bibitem{doi:10.1146/annurev.physchem.49.1.1}
J.~P. Toennies and A.~F. Vilesov.
\newblock \enquote{Spectroscopy of atoms and molecules in liquid helium}.
\newblock Annu. Rev. Phys. Chem., \textbf{49(1)}, 1--41 (1998).

\bibitem{doi:10.1063/1.5028359}
A.~S. Chatterley, E.~T. Karamatskos, C.~Schouder, L.~Christiansen, A.~V.
  Jorgensen, T.~Mullins, J.~Küpper, and H.~Stapelfeldt.
\newblock \enquote{Communication: Switched wave packets with spectrally
  truncated chirped pulses}.
\newblock J. Chem. Phys., \textbf{148(22)}, 221105 (2018).

\bibitem{rapacioli06}
M.~Rapacioli, F.~Calvo, C.~Joblin, P.~Parneix, D.~Toublanc, and F.~Spiegelman.
\newblock \enquote{Formation and destruction of polycyclic aromatic hydrocarbon
  clusters in the interstellar medium}.
\newblock A \& A, \textbf{460}, 519--531 (2006).

\bibitem{vanoanh02}
N.~T. Van~Oanh, P.~Parneix, and P.~Br\'echignac.
\newblock \enquote{Vibrational dynamics of the neutral naphthalene molecule
  from a tight-binding approach}.
\newblock J. Phys. Chem. A, \textbf{106}, 10144--10151 (2002).

\bibitem{opls}
W.~L. Jorgensen, D.~S. Maxwell, and J.~Tirado-Rives.
\newblock \enquote{Development and testing of the \uppercase{OPLS} all-atom
  force field on conformational energetics and properties of organic liquids}.
\newblock J. Am. Chem. Soc., \textbf{118}, 11225--11236 (1996).

\bibitem{vdw83}
B.~van~de Waal.
\newblock \enquote{Calculated ground-state structures of 13-molecule clusters
  of carbon dioxide, methane, benzene, cyclohexane, and naphthalene}.
\newblock J. Chem. Phys., \textbf{79}, 3948--3961 (1983).

\bibitem{mortier06}
W.~J. Mortier, S.~K. Ghosh, and S.~Shankar.
\newblock \enquote{Electronegativity-equalization method for the calculation of
  atomic charges in molecules}.
\newblock J. Am. Chem. Soc., \textbf{108}, 4315--4320 (1986).

\bibitem{rapacioli05}
M.~Rapacioli, F.~Calvo, F.~Spiegelman, C.~Joblin, and D.~J. Wales.
\newblock \enquote{Stacked clusters of polycyclic aromatic hydrocarbon
  molecules}.
\newblock J. Phys. Chem. A, \textbf{109}, 2487--2497 (2005).

\bibitem{G09}
M.~J. Frisch, G.~W. Trucks, H.~B. Schlegel, G.~E. Scuseria, M.~A. Robb, J.~R.
  Cheeseman, G.~Scalmani, V.~Barone, B.~Mennucci, G.~A. Petersson, et~al.
\newblock \enquote{Gaussian~09 {R}evision {D}.01} (2009).
\newblock Gaussian Inc. Wallingford CT.

\bibitem{Chatterley2019}
A.~S. Chatterley, C.~Schouder, L.~Christiansen, B.~Shepperson, M.~H. Rasmussen,
  and H.~Stapelfeldt.
\newblock \enquote{Long-lasting field-free alignment of large molecules inside
  helium nanodroplets}.
\newblock Nat. Commun., \textbf{10(1)}, 133 (2019).

\bibitem{larsen_three_2000}
J.~J. Larsen, K.~Hald, N.~Bjerre, H.~Stapelfeldt, and T.~Seideman.
\newblock \enquote{Three {Dimensional} {Alignment} of {Molecules} {Using}
  {Elliptically} {Polarized} {Laser} {Fields}}.
\newblock Phys. Rev. Lett., \textbf{85(12)}, 2470 (2000).

\bibitem{nevo_laser-induced_2009-1}
I.~Nevo, L.~Holmegaard, J.~H. Nielsen, J.~L. Hansen, H.~Stapelfeldt,
  F.~Filsinger, G.~Meijer, and J.~K{\"u}pper.
\newblock \enquote{Laser-induced 3\uppercase{D} alignment and orientation of
  quantum state-selected molecules}.
\newblock Phys. Chem. Chem. Phys., \textbf{11(42)}, 9912--9918 (2009).

\bibitem{nauta_molecular_1999}
K.~Nauta, D.~T. Moore, and R.~E. Miller.
\newblock \enquote{Molecular orientation in superfluid liquid helium droplets:
  high resolution infrared spectroscopy as a probe of solvent-solute
  interactions}.
\newblock Faraday Discuss., \textbf{113}, 261--278 (1999).

\bibitem{kjeldsen_influence_2005}
T.~K. Kjeldsen, C.~Z. Bisgaard, L.~B. Madsen, and H.~Stapelfeldt.
\newblock \enquote{Influence of molecular symmetry on strong-field ionization:
  {Studies} on ethylene, benzene, fluorobenzene, and chlorofluorobenzene}.
\newblock Phys. Rev. A, \textbf{71(1)}, 013418--12 (2005).

\bibitem{petretti_alignment-dependent_2010}
S.~Petretti, Y.~V. Vanne, A.~Saenz, A.~Castro, and P.~Decleva.
\newblock \enquote{Alignment-dependent ionization of \uppercase{N}$_{2}$,
  \uppercase{O}$_{2}$, and \uppercase{CO}$_{2}$ in intense laser fields}.
\newblock Phys. Rev. Lett., \textbf{104}, 223001 (2010).

\bibitem{hansen_orientation-dependent_2012}
J.~L. Hansen, L.~Holmegaard, J.~H. Nielsen, H.~Stapelfeldt, D.~Dimitrovski, and
  L.~B. Madsen.
\newblock \enquote{Orientation-dependent ionization yields from strong-field
  ionization of fixed-in-space linear and asymmetric top molecules}.
\newblock J. Phys. B: At. Mol. Opt. Phys., \textbf{45(1)}, 015101 (2012).

\bibitem{mikosch_channel-_2013}
J.~Mikosch, A.~E. Boguslavskiy, I.~Wilkinson, M.~Spanner, S.~Patchkovskii, and
  A.~Stolow.
\newblock \enquote{Channel- and {Angle}-{Resolved} {Above} {Threshold}
  {Ionization} in the {Molecular} {Frame}}.
\newblock Phys. Rev. Lett., \textbf{110(2)}, 023004 (2013).

\bibitem{boll_imaging_2014}
R.~Boll, A.~Rouzée, M.~Adolph, D.~Anielski, A.~Aquila, S.~Bari, C.~Bomme,
  C.~Bostedt, J.~D. Bozek, H.~N. Chapman, L.~Christensen, R.~Coffee,
  N.~Coppola, S.~De, P.~Decleva, S.~W. Epp, B.~Erk, F.~Filsinger, L.~Foucar,
  T.~Gorkhover, L.~Gumprecht, A.~Hömke, L.~Holmegaard, P.~Johnsson, J.~S.
  Kienitz, T.~Kierspel, F.~Krasniqi, K.-U. Kühnel, J.~Maurer,
  M.~Messerschmidt, R.~Moshammer, N.~L.~M. Müller, B.~Rudek, E.~Savelyev,
  I.~Schlichting, C.~Schmidt, F.~Scholz, S.~Schorb, J.~Schulz, J.~Seltmann,
  M.~Stener, S.~Stern, S.~Techert, J.~Thøgersen, S.~Trippel, J.~Viefhaus,
  M.~Vrakking, H.~Stapelfeldt, J.~Küpper, J.~Ullrich, A.~Rudenko, and
  D.~Rolles.
\newblock \enquote{Imaging molecular structure through femtosecond
  photoelectron diffraction on aligned and oriented gas-phase molecules}.
\newblock Faraday Discuss., \textbf{171}, 57--80 (2014).

\bibitem{johansen_generation_2016}
R.~Johansen.
\newblock {PhD} {Thesis}, Aarhus University (2016).

\bibitem{heidenreich01}
A.~Heidenreich, U.~Even, and J.~Jortner.
\newblock \enquote{Nonrigidity, delocalization, spatial confinement and
  electronic-vibrational spectroscopy of anthracene-helium clusters}.
\newblock J. Chem. Phys., \textbf{115}, 10175--10185 (2001).

\bibitem{calvo12}
F.~Calvo.
\newblock \enquote{Size-induced melting and reentrant freezing in
  fullerene-doped helium clusters}.
\newblock Phys. Rev. B, \textbf{85}, 060502(R) (2012).

\bibitem{calvoyb16}
F.~Calvo, E.~Yurtsever, and {\"O}.~Birer.
\newblock \enquote{Possible formation of metastable pah dimers upon pickup by
  helium droplets}.
\newblock J. Phys. Chem. A, \textbf{120}, 1717--1736 (2016).

\bibitem{christensen_dynamic_2014}
L.~Christensen, J.~H. Nielsen, C.~B. Brandt, C.~B. Madsen, L.~B. Madsen, C.~S.
  Slater, A.~Lauer, M.~Brouard, M.~P. Johansson, B.~Shepperson, and
  H.~Stapelfeldt.
\newblock \enquote{Dynamic {Stark} {Control} of {Torsional} {Motion} by a
  {Pair} of {Laser} {Pulses}}.
\newblock Phys. Rev. Lett., \textbf{113(7)}, 073005 (2014).

\bibitem{B810189B}
J.-D. Chai and M.~Head-Gordon.
\newblock \enquote{Long-range corrected hybrid density functionals with damped
  atom-atom dispersion corrections}.
\newblock Phys. Chem. Chem. Phys., \textbf{10}, 6615--6620 (2008).

\bibitem{doi:10.1021/ct401026a}
F.~Jensen.
\newblock \enquote{Unifying general and segmented contracted basis sets.
  segmented polarization consistent basis sets}.
\newblock J. Chem. Theory Comput., \textbf{10(3)}, 1074--1085 (2014).

\bibitem{doi:10.1063/1.1433747}
L.~Jensen, P.-O. Åstrand, A.~Osted, J.~Kongsted, and K.~V. Mikkelsen.
\newblock \enquote{Polarizability of molecular clusters as calculated by a
  dipole interaction model}.
\newblock J. Chem. Phys., \textbf{116(10)}, 4001--4010 (2002).

\bibitem{Atkins}
P.~Atkins and R.~Friedman.
\newblock \emph{Molecular Quantum Mechanics, 4th Edition} (Oxford University
  press, Oxford, 2005).

\bibitem{lapack99}
D.~Manura and D.~Dahl.
\newblock \emph{{SIMION (R) 8.0} User Manual} (Scientific Instrument Services,
  Inc. Ringoes, NJ 08551, 2008).

\bibitem{doi:10.1063/1.1147588}
R.~M. Wood, Q.~Zheng, A.~K. Edwards, and M.~A. Mangan.
\newblock \enquote{Limitations of the axial recoil approximation in
  measurements of molecular dissociation}.
\newblock Rev. Sci. Instrum., \textbf{68(3)}, 1382--1386 (1997).

\bibitem{Lauge}
L.~Christensen, L.~Christiansen, B.~Shepperson, and H.~Stapelfeldt.
\newblock \enquote{Deconvoluting nonaxial recoil in coulomb explosion
  measurements of molecular axis alignment}.
\newblock Phys. Rev. A, \textbf{94}, 023410 (2016).

\end{thebibliography}
\end{document}